\catcode`\@=11

\newif\if@fewtab\@fewtabtrue

{\count255=\time\divide\count255 by 60
\xdef\hourmin{\number\count255}
\multiply\count255 by-60\advance\count255 by\time
\xdef\hourmin{\hourmin:\ifnum\count255<10 0\fi\the\count255}}
\def\ps@draft{\let\@mkboth\@gobbletwo
    \def\@oddhead{}
    \def\@oddfoot
       {\hbox to 7 cm{$\scriptstyle Draft\ version:\ \draftdate$
       \hfil}\hskip -7cm\hfil\rm\thepage \hfil}
    \def\@evenhead{}\let\@evenfoot\@oddfoot}

\def\draftcite#1{\ifnum\draftcontrol=1#1\else{}\fi}

\def\@lbibitem[#1]#2{\item{}\hskip -3cm \hbox to 2cm
{\hfil$\scriptstyle\draftcite{#2}$}\hskip
1cm[\@biblabel{#1}]\if@filesw
     {\def\protect##1{\string ##1\space}\immediate
      \write\@auxout{\string\bibcite{#2}{#1}}}\fi\ignorespaces}

\def\@bibitem#1{\item\hskip -3cm \hbox to 2cm
{\hfil $\scriptstyle\draftcite{#1}$}\hskip 1cm
\if@filesw \immediate\write\@auxout
       {\string\bibcite{#1}{\the\value{\@listctr}}}\fi\ignorespaces}

\def\nsection#1{\section{#1}\setcounter{equation}{0}}

\font\tendl=msbm10  scaled \magstep1
\font\sevendl=msbm7 scaled \magstep1
\font\fivedl=msbm5 scaled \magstep1
\font\tengl=eufm10  scaled \magstep1
\font\sevengl=eufm7 scaled \magstep1
\font\fivegl=eufm5 scaled \magstep1

\newfam\dlfam  
\textfont\dlfam=\tendl \scriptfont\dlfam=\sevendl
\scriptscriptfont\dlfam=\fivedl
\newfam\glfam  
\textfont\glfam=\tengl \scriptfont\glfam=\sevengl
\scriptscriptfont\glfam=\fivegl

\def\draftdate{\number\month/\number\day/\number\year\ \ \ \hourmin }

\global\def\draftcontrol{0}
\catcode`\@=12
\documentclass[aps,superscriptaddress,floatfix]{revtex4}
\usepackage{epsfig}
\usepackage[centerlast]{subfigure}
\usepackage{bm}
\usepackage{amssymb}
\usepackage{verbatim}
\usepackage{graphicx}

\newcommand{\be}{\begin{eqnarray}}
\newcommand{\en}{\end{eqnarray}\vs 0.5 cm}

\newcommand{\vs}{\vskip}

\newcommand{\ff}{{\hspace{-0.02cm}f}}
\newcommand{\ii}{{i}}

\newcommand{\qq}{\begin{eqnarray}}

\newcommand{\qqq}{\end{eqnarray}}

\newcommand{\CA}{{\cal A}}

\newcommand{\CK}{{\cal K}}

\begin{document}

\
\vskip 1cm
\centerline{\Large{\bf{Refined}}}
\vskip 0.22cm
\centerline{\Large{\bf{Second \,Law \,of 
 \,Thermodynamics}}}
\vskip 0.1cm
\centerline{\Large{\bf{for \,fast \,random \,processes}}}
\vskip 0.8cm
\centerline{Erik Aurell$^{1,2,3}$,\,\ \ Krzysztof Gaw\c{e}dzki$^{4}$,\,\ \ 
Carlos Mej\'{\i}a-Monasterio$^{5}$,}
\vskip 0.1cm
\centerline{Roya Mohayaee$^{6}$,\,\ \ Paolo Muratore-Ginanneschi$^{7}$}
\vskip 0.7cm
\centerline{$^{1}$\it\,ACCESS Linnaeus Center, KTH, 10044 Stockholm, Sweden}
\vskip 0.25cm
\centerline{$^{2}$\it\,Computational Biology Department, KTH,}
\centerline{\it AlbaNova University Center, 10691 Stockholm, Sweden}
\vskip 0.25cm
\centerline{$^{3}$\it\,Department of Information and Computer Science,
Aalto University,} 
\centerline{\it PO Box 15400, 00076 Aalto, Finland}
\vskip 0.25cm
\centerline{$^{4}$\it\,CNRS, Laboratoire de Physique, ENS Lyon, Universit\'e 
de Lyon,}
\centerline{\it 46 All\'ee d'Italie, 69364 Lyon, France}
\vskip 0.25cm
\centerline{$^{5}$\it\,Laboratory of Physical Properties,} 
\centerline{\it Department of Rural Engineering, Technical University of Madrid,} 
\centerline{\it Avenida Complutense s/n, 28040 Madrid, Spain}
\vskip 0.25cm
\centerline{$^{6}$\it\,CNRS, Institut d'Astrophysique de Paris,
Universit\'e Pierre et Marie Curie,} 
\centerline{\it8 bis boulevard Arago, 75014 Paris, France} 
\vskip 0.25cm
\centerline{$^{7}$\it\,Department of Mathematics
and Statistics, University of Helsinki,} 
\centerline{\it P.O. Box 68 FIN-00014, Helsinki, Finland}
\vskip 0.8cm

\centerline{\bf Abstract}
\vskip 0.2cm

\noindent{We establish a refined version of the Second Law of
Thermodynamics for Langevin stochastic processes describing  
mesoscopic systems driven by conservative or non-conservative forces 
and interacting with thermal noise. The refinement is based 
on the Monge-Kantorovich optimal mass transport. General 
discussion is illustrated by numerical analysis of a model 
for micron-size particle manipulated by optical tweezers.}

\date{ }

\vskip 1cm

\nsection{Introduction}

\noindent In recent years an increased interest in fluctuations of 
mesoscopic systems interacting with noisy environment has led to 
the development of ``Stochastic Thermodynamics'' that revisited
relations between thermodynamical principles and statistical 
description within simple models based on stochastic differential 
equations, see \citep{Seifert,Sekim} and references
therein. The aim of this note is to make a junction between two circles 
of ideas in the context of Stochastic Thermodynamics of systems whose 
evolution is described by the overdamped Langevin equation. One circle 
concerns the versions \citep{HaSa,DLutz} of the Second Law 
of Thermodynamics and of its reformulation in the framework of Thermodynamics 
of Computation by Landauer \citep{Landauer,Bennett}. The other circle deals 
with the optimal control problems in Stochastic Thermodynamics that were 
recently connected in \citep{AMMMG} to the Monge-Kantorovich optimal mass 
transport and the Burgers equation. The result of the junction will 
be a refinement, relevant for fast processes, of the Second Law of 
Stochastic Thermodynamics. Our improvement of the Second Law does not go
in the direction of a better control of fluctuations of thermodynamical 
quantities \citep{Szil}, as do various Fluctuation Relations studied 
intensively in last years, see \citep{EVS,Gall,Jarzyn1}. Instead, 
it establishes the optimal lower bound on the total entropy production 
in non-equilibrium processes of fixed duration.
\vskip 0.1cm

The paper is organized as follows. In Sec.\,\ref{sec:secondLaw}, we
define for overdamped Langevin evolution with conservative driving forces
the concepts of performed work, heat release, and entropy production, and we
recall the basic laws of Stochastic Thermodynamics. Sec.\,\ref{sec:Landauer} 
contains a brief discussion of the relation between the Second Law of 
Thermodynamics and the Landauer principle. In Sec.\,\ref{sec:optcontr}, 
we replace the minimization of the total entropy production in overdamped
Langevin processes that interpolate in a fixed time window between given 
statistical states by a minimization problem considered by 
Benamou-Brenier in \citep{BB} and shown there to be equivalent to 
the Monge-Kantorovich optimal mass transport problem that is the subject 
of Sec.\,\ref{sec:MK}. The latter two sections briefly review 
the classical mathematical results about the optimal mass transport \citep{V} 
needed in our argument. In particular, the approach of \citep{BB} establishes 
a direct connection between the Monge-Kantorovich problem and the inviscid 
Burgers equation for potential velocities that plays a crucial role below. 
On the basis of the above results, we establish in Sec.\,\ref{sec:R2ndLaw} 
the refined version of the Second Law of Stochastic 
Thermodynamics, applying it to the example of Gaussian processes 
considered already in a similar context in \citep{AMMMG} and, in a special 
case, in \citep{SS}. Sec.\,\ref{sec:RLandauer} discusses the corresponding 
refinement of the Landauer principle, illustrating it by the numerical 
analysis of a simple model of a micron-size particle in time-dependent 
optical traps. Sec.\,\ref{sec:NCF} extends the refined Second Law to 
the case of Langevin evolutions with non-conservative forces, showing 
that the preceding analysis covers also that case. Conclusions 
and remarks about open problems make up Sec.\,\ref{sec:conc}.
\vskip 0.3cm

\noindent{\bf Acknowledgements.} \ K.G. thanks S. Ciliberto and U. Seifert
for inspiring discussions. His work was partly done within the 
framework of the STOSYMAP project ANR-11-BS01-015-02. R.M.'s work was partly 
supported by the OTARIE project ANR-07-BLAN-0235. P.M.-G. acknowledges
support of the Center of Excellence "Analysis and Dynamics" of the 
Academy of Finland.

\nsection{Stochastic Thermodynamics for Langevin equation}
\label{sec:secondLaw}

\noindent We consider small statistical-mechanical systems, for example
composed of mesoscopic particles, driven by time-dependent conservative
forces and interacting with a noisy environment. The temporal evolution 
of such a system my often be well described by the overdamped stochastic 
Langevin equation
\qq
d\bm x\ =\ -{M}\hspace{0.02cm}\bm\nabla U(t,\bm x)\,dt\,
+\,d\bm\zeta(t)
\label{Lan}
\qqq
in $\,d$-dimensional space of configurations, with a smooth potential 
$\,U(t,\bm x)\,$ and a white noise $\,d\bm\zeta(t)\,$ whose covariance 
is 
\qq
\big\langle d\zeta^a(t)\,d\zeta^b(t')\big\rangle\,
=\,2\hspace{0.03cm}D^{ab}\,\delta(t-t')\,dt\,,
\qqq
where $\,\big\langle\,-\,\big\rangle\,$ denotes the expectation value.
The mobility and diffusivity matrices $\,{M}=({M}^{ab})\,$ and 
$\,D=({D}^{ab})\,$ occurring above are assumed positive and 
$\bm x$-independent (the latter assumption is for the sake of simplicity 
and could be 
relaxed at the cost of few corrective terms). To assure that 
the noise models the thermal environment at absolute 
temperature $\,T$, \,we impose the Einstein relation
\qq
D\,=\,k_BT\,{M}\,,
\label{ER}
\qqq
where $\,k_B\,$ is the Boltzmann constant. \,Potentials $\,U_t(\bm x)\equiv 
U(t,\bm x)\,$ are assumed to be sufficiently confining so that 
the solutions of the stochastic equation (\ref{Lan}) do not explode 
in finite time. Given a probability density $\,\rho_\ii(\bm x)\,$ 
at the initial time $\,t=0$, \,they define then for $\,t\geq 0\,$ a, 
in general non-stationary, Markov diffusion 
process $\,\bm x(t)$. The relation
\qq
\frac{_d}{^{dt}}\,\big\langle\,g(\bm x(t)\,\big\rangle\ =\ 
\big\langle\,(L_t g)(\bm x(t)\,\big\rangle\,,
\qqq
holding for smooth functions $\,g(\bm x)$, \,determines 
the $2^{\rm nd}$ order differential operator
\qq
L_t\ =\ -(\bm\nabla U_t)\cdot{M}\,\bm\nabla
+k_BT\hspace{0.07cm}\bm\nabla\cdot{M}\,\bm\nabla\,,
\label{Lt}
\qqq
the (time-dependent) generator of diffusion $\,\bm x(t)$. 
\,The instantaneous distributions of the 
process, describing its statistical properties at fixed times, are given 
by the probability densities
\qq
\rho(t,\bm x)\ =\ \big\langle\,\delta(\bm x-\bm x(t))\,\big\rangle\,
\equiv\,\exp\Big[-\frac{_{R(t,\bm x)}}{^{k_BT}}\Big]\,,
\label{Rt}
\qqq
that we assume smooth, positive, and with finite moments.
They evolve according to the Fokker-Planck equation
\qq
\partial_t\rho\,=\,L_t^\dagger\rho\,,
\label{rhoev}
\qqq
where $\,L_t^\dagger\,$ is the $2^{\rm nd}$ order differential operator
adjoint to $\,L_t\,$. \,Explicitly,
\qq
L_t^\dagger\ =\ {\bm\nabla}\cdot M\hspace{0.03cm}({\bm\nabla}U_t)\,+\,k_B\,
T\,{\bm\nabla}\cdot M\,{\bm\nabla}\,.
\label{Ldagger}
\qqq
In what follows, it will be crucial that the Fokker-Planck equation 
(\ref{rhoev}) may be rewritten as the advection equation
\qq
\partial_t\rho+\bm\nabla\cdot(\rho\,{\bm v})\,=\,0
\label{rhoadv}
\qqq
in the deterministic velocity field 
\qq
\bm v(t,\bm x)\,&=&\,-M\,\big({\bm\nabla}U\,+\,k_BT\,\rho^{-1}
{\bm\nabla}\rho\big)(t,\bm x)\ =\ -M\,{\bm\nabla}(U-R)(t,\bm x)\,.
\label{mlv}
\qqq
The time-dependent vector field $\,\bm v(t,\bm x)$, \,called
current velocity in \citep{Nelson}, has the interpretation of the mean 
local velocity of the process $\,\bm x(t)\,$ defined by the limiting 
procedure 
\qq
\bm v(t,\bm x)\ =\ \lim\limits_{\epsilon\to0}\,\frac{\big\langle\,
\delta(\bm x-\bm x(t))\,(\bm x(t+\epsilon)-\bm x(t-\epsilon))\,\big\rangle}
{2\hspace{0.02cm}\epsilon\,\big\langle\,\delta(\bm x-\bm x(t))\,\big\rangle}
\label{limpr}
\qqq
(the limit has to be taken after the expectation as the trajectories
of the diffusion process are not differentiable).
\vskip 0.3cm

The setup of Langevin equation permits simple definitions of thermodynamical 
quantities. The fluctuating (i.e. trajectory-dependent) work performed on 
the system between initial time $\,t=0\,$ and final time $\,t=t_\ff\,$ is 
given by the Jarzynski expression
\citep{Jarzyn}
\qq
W\ =\ \int_{0}^{t_\ff}\hspace{-0.1cm}\partial_tU(t,\bm x(t))\hspace{0.08cm}dt
\label{DW}
\qqq
and the fluctuating heat released into the environment during the same
time interval by the formula
\qq
Q\ =\ -\int_{0}^{t_\ff}\hspace{-0.05cm}{\bm\nabla}
U(t,\bm x(t))\cdot\circ\,d\bm x(t)
\label{DQ}
\qqq
with the Stratonovich stochastic integral (symbolized by \,''$\circ$'').
The expectation value of work is then given by the identity
\qq
\big\langle\,W\,\big\rangle&\,=\,&\int_{0}^{t_\ff}\hspace{-0.1cm}dt
\int\partial_tU(t,\bm x)\,\rho(t,\bm x)\,dx\,,\label{avDU}
\qqq
where $\,dx\,$ denotes the standard $\,d$-dimensional volume element.
In order to calculate the expectation value of heat release,
one rewrites the definition (\ref{DQ}) in terms of the It\^{o} stochastic 
integral, including a corrective term:
\qq
Q\,&=&\,\,-\int_{0}^{t_\ff}\hspace{-0.05cm}{\bm\nabla}U(t,\bm x(t))
\cdot d\bm x(t)
\,-\,k_BT
\int_{0}^{t_\ff}\hspace{-0.05cm}{\bm\nabla}\cdot M\,{\bm\nabla}U(t,\bm x(t))\,\,
dt\cr
&=&\hspace{-0.05cm}\int_{0}^{t_\ff}\hspace{-0.1cm}\big(({\bm\nabla}U)
\cdot{M}\,({\bm\nabla}U)
\,-\,k_BT\,({\bm\nabla}\cdot{M}\,{\bm\nabla}U)\big)(t,\bm x(t))\,\,dt
\,-\,\int_{0}^{t_\ff}\hspace{-0.06cm}{\bm\nabla}U(t,\bm x(t))\cdot 
d\bm\zeta(t)\,.\qquad
\qqq
The last term does not contribute to the expectation value due to the 
martingale property of the It\^o integral so that
\qq
\big\langle\,Q\,\big\rangle\ =\ 
\int_{0}^{t_\ff}\hspace{-0.1cm}dt\int
\big(({\bm\nabla}U)\cdot{M}\,({\bm\nabla}U)
\,-\,k_BT\,({\bm\nabla}\cdot{M}\,{\bm\nabla}U)\big)(t,\bm x)\,
\,\rho(t,\bm x)\,\,dx\,,
\label{avQ}
\qqq
or, \,upon integration by parts over space,
\qq
\big\langle\,Q\,\big\rangle\ =\ 
\int_{0}^{t_\ff}\hspace{-0.1cm}dt\int
\big(\bm\nabla U-\bm\nabla R\big)(t,\bm x)\cdot{M}\,({\bm\nabla}U)(t,\bm x)\,
\,\rho(t,\bm x)\,\,dx
\label{avQ1}
\qqq
(here and below, we assume that the spatial boundary terms in integration 
by parts vanish; this is assured for confining potentials and fast decaying
initial density of the process).
\vskip 0.3cm

\,The First Law of Thermodynamics, expressing the conservation of energy, 
takes in the context of overdamped Langevin dynamics the form of an identity 
\qq
W\,-\,Q\ =\ \Delta U
\label{1stLawfl}
\qqq
\vskip -0.3cm
\noindent where
\vskip -0.3cm
\qq
\Delta U\,=\,U(t_\ff,\bm x(t_\ff))\,-\,U(0,\bm x(0))
\label{DU}
\qqq
\vskip 0.2cm
\noindent is the difference of the potential energy between the end point
and the initial point of the process trajectory. Eq.\,(\ref{1stLaw}) holds 
for the fluctuating quantities and not only as the relation
\qq
\big\langle\,W\,\big\rangle\,-\,\big\langle\,Q\,\big\rangle\ =\ 
\big\langle\,\Delta U\,\big\rangle
\label{1stLaw}
\qqq
for the expectation values with
\qq
\big\langle\,\Delta U\,\big\rangle&\,=\,&\int U(t_\ff,\bm x)\,\,
\rho_\ff(\bm x)\,\,dx\,-\,\int U(0,\bm x)\,\,
\rho_\ii(\bm x)\,\,dx\,,\label{avW}
\qqq
where $\,\rho_\ii\equiv\rho_0\,$ and $\,\rho_\ff\equiv\rho_{t_\ff}$.
\vskip 0.2cm

In \citep{SS} and \citep{AMMMG} it was assumed that at the initial and 
the final times the potential may undergo jumps from $\,U_\ii(\bm x)\,$ 
to $\,U(0,\bm x)\,$ and from $\,U(t_\ff,\bm x)\,$ to $\,U_\ff(\bm x)$, 
\,leading to the modified expression for the work
\qq
W\ =\ U(0,\bm x(0))-U_\ii(\bm x(0))\,+\int_{0}^{t_\ff}\hspace{-0.1cm}
\partial_tU(t,\bm x(t))\hspace{0.08cm}dt\,+\,U_\ff(\bm x(t_\ff))
-U(t_\ff,\bm x(t_\ff))\,,
\label{DW1}
\qqq
with the contributions from the initial and final jumps of the potential 
included. Of course, Eq.\,(\ref{DW1}) may
be obtained from (\ref{DW}) by an appropriate limiting procedure
where the jumps are smoothened over short initial and final time
intervals. Within such a procedure, the process itself is not modified
in the limit on the time interval $\,[0,t_\ff]\,$ and the limiting heat 
release is still given by expression 
(\ref{DQ}). The First Law (\ref{1stLawfl}) continues to hold, provided 
we replace formula (\ref{DU}) for $\,\Delta U\,$ by
\qq
\Delta U\,=\,U_\ff(\bm x(t_\ff))\,-\,U_\ii(\bm x(0))\,.
\label{DU1}
\qqq
The expectation value of work may now be expressed in the form
\qq
\big\langle\,W\,\big\rangle\ =\ \int U_\ff(\bm x)\,\,\rho_\ff(\bm x)\,\,dx\,-\,\int U_\ii(\bm x)\,\,\rho_\ii(\bm x)\,\,dx\ +\ \big\langle\,Q\,
\big\rangle
\label{avW1}
\qqq
with the average heat release given by Eq.\,(\ref{avQ1}).
\vskip 0.3cm

Let us pass to the discussion of the Second Law of Thermodynamics in the 
context of Langevin dynamics (\ref{Lan}) (eventual jumps of potential 
at the ends of the time interval will not affect the formulae below). 
The instantaneous entropy of the system is given by the usual 
Gibbs-Shannon formula
\qq
S_{sys}(t)\ =\ -\,k_B\int\ln(\rho(t,\bm x))\,\,\rho(t,\bm x)\,\,dx\,.
\label{entropy}
\qqq
For its time derivative, one obtains from the Fokker-Planck equation
(\ref{rhoev}) the expression
\qq
\frac{d}{dt}S_{sys}(t)\,&=&\,-k_B\int\ln(\rho_t(\bm x))\,
(L_t^\dagger\rho_t)(\bm x)\,dx\cr\cr
&=&\,\frac{1}{T}\int R_t(\bm x)\,\big({\bm\nabla}\cdot
M\,({\bm\nabla}U_t)\,
-\,{\bm\nabla}\cdot M\,({\bm\nabla}R_t)\big)(\bm x)\,\,\rho_t(\bm x)\,\,
dx\cr\cr
&=&\,-\,\frac{1}{T}\int({\bm\nabla}R_t)(\bm x)\cdot M\,
(\bm\nabla U_t-\bm\nabla R_t)(\bm x)\,\,\rho_t(\bm x)\,\,dx\,,
\label{chSsys}
\qqq 
where the last equality follows again by integration by parts.
Integrating over time, one gets for 
the change of the entropy of the system in the time interval 
$\,[0,t_\ff]\,$ the formula
\qq
\Delta S_{sys}\,\equiv\,S_{sys}(t_\ff)-S_{sys}(0)
\ =\ -\,\frac{1}{T}\int_{0}^{t_\ff}\hspace{-0.1cm}dt
\int{(\bm\nabla}R)(t,\bm x)\cdot M\,(\bm\nabla U-\bm\nabla R)(t,\bm x)
\,\,\rho(t,\bm x)\,\,dx\,.\quad
\label{DSsys}
\qqq
Since the system evolves interacting with the thermal environment,
the entropy of the latter also changes. The change of entropy 
of environment is related to the average heat release by the thermodynamic
formula
\qq
\Delta S_{env}\ =\ \frac{1}{T}\,\big\langle\,Q\,\big\rangle\,.
\label{DSenv}
\qqq
For the total entropy production, Eqs.\,(\ref{avQ1}) and (\ref{DSsys})
give
\qq
&&\cr
\Delta S_{tot}\,&=&\,\Delta S_{sys}\,+\,\Delta S_{env}\cr\cr
&=&\,\frac{1}{T}\int_{0}^{t_\ff}\hspace{-0.1cm}dt\int
({\bm\nabla}U-\bm\nabla R)(t,\bm x)\cdot M\,(\bm\nabla U-\bm\nabla R)
(t,\bm x)\,\,\rho(t,\bm x)\,\,dx\cr
&=&\,\frac{1}{T}\int_{0}^{t_\ff}\hspace{-0.1cm}dt\int
\big(\bm v\cdot M^{-1}\bm v\big)(t,\bm x)\,\,\rho(t,\bm x)\,\,dx\,,
\label{DStot}
\qqq
where the last line is expressed in terms of the mean local velocity
given by Eq.\,(\ref{mlv}). \,Similar formulae for the entropy
production appeared e.g. in \citep{Seifert1,MaesNW,FR,AMMMG1}. 
In the obvious way, identity (\ref{DStot}) implies the Second Law 
of Stochastic Thermodynamics:
\qq
\Delta S_{tot}\ \geq\ 0
\label{2ndLaw}
\qqq
stating that the total entropy production composed of the changes
of entropy of the system and of the environment has to be non-negative. 
Inequality (\ref{2ndLaw}) may be also rewritten as a lower bound 
for the average heat release:
\qq
\big\langle\,Q\,\big\rangle\,\geq\,-T\,\Delta S_{sys}\,.
\label{bdforQ}
\qqq

\nsection{Landauer principle}
\label{sec:Landauer}

\noindent 
In the form (\ref{bdforQ}), the Second Law of Stochastic 
Thermodynamics is closely related to the Landauer principle 
\citep{Landauer,Bennett} stating that the erasure of one bit of 
information during a computation process conducted in thermal 
environment requires a release of heat equal (in average) to at 
least $\,(\ln{\hspace{-0.04cm}2})\hspace{0.03cm}k_BT$. 
\,As an example, consider a bi-stable 
system that may be in two distinct states and undergoes a process 
that at final time leaves it always in, say, the second of those states. 
Such a device may be realized in the context of Stochastic Thermodynamics 
by an appropriately designed Langevin evolution that starts from 
the Gibbs state corresponding to a potential with two symmetric 
wells separated by a high barrier and ends in a Gibbs state 
corresponding to a potential with only one of those wells 
\citep{DLutz}. The change of system entropy in such a process is 
approximately 
\qq
\Delta S_{sys}\,=\,-\hspace{0.03cm}\,(\ln{\hspace{-0.05cm}1})\hspace{0.03cm}k_B\,
+\,2\hspace{0.03cm}(\ln{\hspace{-0.04cm}\frac{_1}{^2}})\hspace{0.03cm}
\frac{_1}{^2}\hspace{0.03cm}k_B\,
=\,-\hspace{0.03cm}(\ln{\hspace{-0.04cm}2})\hspace{0.03cm}k_B\,
\qqq
and Landauer's lower bound for average heat release follows from
inequality (\ref{bdforQ}). Note that in this situation we fix the initial 
and the final state of Langevin evolution, inquiring how much heat 
is released during a process that interpolates between those states. 
As is well known, in order to saturate the lower bounds (\ref{2ndLaw}) 
or (\ref{bdforQ}), one has to move infinitely slowly so that the system 
passes at intermediate times through a sequence of equilibrium states. 
Suppose however, that we cannot afford to go too slowly. Indeed, in 
computational devices, we are interested in fast dynamics that arrives 
at the final state quickly but produces as little heat as possible. We 
are therefore naturally led to two questions:
\begin{itemize}
\item What is the lower bound for the total entropy production or the 
average heat release in the process that interpolates between given 
states in a time interval of fixed length?
\item What is the dynamical protocol that leads to such a minimal
total entropy production or heat release?
\end{itemize}
These questions make sense in more general setups but we shall
study them below in the context of Stochastic Thermodynamics  
of processes described by Langevin equation (\ref{Lan}).
The initial and final states will be given by probability
densities $\,\rho_\ii(\bm x)\,$ and $\,\rho_\ff(\bm x)$. \,The dynamical 
protocols will be determined by specifying for $\,0\leq t\leq t_\ff$ a 
time dependent steering potential $\,U(t,\bm x)$, that will be called 
the ``control'' below. \,In such a setup, the question about the minimum 
of total entropy production or average heat release becomes an optimization 
problem in Control Theory \citep{GM,FlemSon}. It was recently discussed, 
together with the optimization of average performed work, in 
refs.\,\citep{SS,AMMMG}, see also 
\citep{GMSS,AMMMG1}.

\nsection{Optimal control of entropy production}
\label{sec:optcontr}

\noindent We shall describe below a relation of the minimization problem 
for total entropy production or the average heat release to the optimal
mass transport \citep{V} and the inviscid Burgers dynamics \citep{Burg}. 
To our knowledge, such a relation was first established in ref.\,\citep{AMMMG} 
using stochastic optimization. Nevertheless, connections between 
stochastic control and (viscous) Burgers equation and between 
Fokker-Planck equation and optimal mass transport are old themes, see e.g. 
Chapter VI of \citep{FlemSon}, or \citep{HSS} in a particular case, for 
the first ones and \citep{JKO} for the second ones. Here, inspired by 
the discussion in \citep{AMMMG1}, we shall minimize the total entropy 
production given by Eq.\,(\ref{DStot}) by a direct argument in the 
spirit of deterministic optimal control.
\vskip 0.2cm

Our strategy is based on the subsequent use of the obvious fact that 
if a minimizer of a function on a bigger set lies in a smaller one 
then it realizes also the minimum of the function over the smaller set. 
We shall minimize the functional 
\qq
\CA[\bm v,\rho_\ii]\ =\ \frac{1}{T}\int_{0}^{t_\ff}\hspace{-0.1cm}dt
\int\big(\bm v\cdot M^{-1}\bm v\big)(t,\bm x)\,\,\rho(t,\bm x)\,\,dx\,,
\label{CS}
\qqq
where $\,\rho(t,\bm x)\,$ is determined by the
advection equation (\ref{rhoadv}) from the initial density 
$\,\rho_\ii\,$ and the velocity field $\,\bm v(t,\bm x)$, \,over all 
velocity fields $\,\bm v\,$ under the constraint that 
$\,\rho(t_\ff,\bm x)=\rho_\ff(\bm x)$. \,Such an extended minimization 
problem was considered in \citep{BB}. The crucial but simple additional
step will be the observation that the optimal velocity field 
$\,\bm v(t,\bm x)\,$ for which the constraint minimum is attained is 
a local mean velocity for a certain control $\,U(t,\bm x)$. \,Such 
an optimal control realizes then the Langevin dynamics that interpolates 
on the time interval $\,[0,t_\ff]\,$ between densities $\,\rho_\ii\,$ 
and $\,\rho_\ff\,$ with minimal total entropy production $\,\Delta S_{tot}$.    
\vskip 0.2cm

In \citep{BB}, see also \citep{BB1}, it was shown how one may reduce 
the constraint minimization of functional (\ref{CS}) to the optimal 
mass transport problem. Here is a slight modification of that argument.
We shall admit smooth velocity fields $\,\bm v\,$ for which the Lagrangian 
trajectories $\,\bm x(t)$ solving the equation
\qq
\dot{\bm x}(t)\,=\,\bm v(t,\bm x(t))\,,
\label{lagr}
\qqq
where the dot stands for $\,t$-derivative,
do not blow up. E.g., we may take $\,\bm v\,$ bounded by a linear 
function of $\,|\bm x|$. \,The solution of the advection equation 
(\ref{rhoadv}) is then given by the formula
\qq
\rho(t,\bm x)\,=\,\int\delta(\bm x-\bm x(t;\bm x_\ii))\,\,
\rho_\ii(\bm x_\ii)\,\,dx_\ii\,,
\label{sFP}
\qqq
where $\,\bm x(t;\bm x_\ii)\,$ denotes the Lagrangian trajectory that
passes through $\,\bm x_\ii\,$ at time $\,t=0$. \,The substitution of
Eq.\,(\ref{sFP}) into definition (\ref{CS}) results in the identity
\qq
&&\CA[\bm v,\rho_\ii]\ =\ \frac{1}{T}
\int_{0}^{t_\ff}\hspace{-0.1cm}dt\int\dot{\bm x}(t;\bm x_\ii)\cdot
M^{-1}\,\dot{\bm x}(t;\bm x_\ii)\,\,\rho_\ii(\bm x_\ii)\,\,dx_\ii\,.
\label{CF}
\qqq
Since velocity field $\,\bm v(t,\bm x)\,$ may be recovered from its 
Lagrangian flow $\,\bm x(t;\bm x_\ii)$, \,the minimization of 
$\,\CA[\bm v,\rho_\ii]\,$ over velocity fields may be replaced by 
the minimization of the right hand side of (\ref{CF}) over 
Lagrangian flows such that the map $\,\bm x_\ii\,\mapsto\,\bm x(t_\ff;
\bm x_\ii)\equiv\bm x_\ff(\bm x_\ii)\,$ is constrained by 
the condition 
\qq
\rho_\ff(\bm x)\,=\,\int\delta(\bm x-\bm x_\ff(\bm x_\ii))\,\,
\rho_\ii(\bm x_\ii)\,\,dx_\ii\,,
\label{crnt0}
\qqq
or, \,equivalently, \,denoting by $\,\frac{\partial\,(\bm x_\ff(\bm x_\ii))}
{\partial(\bm x_\ii)}\,$ the Jacobian of the map $\,\bm x_\ii\mapsto
\bm x_\ff(\bm x_\ii)$, \,by the requirement that
\qq
\rho_\ff(\bm x_\ff(\bm x_\ii))\,\frac{\partial\,(\bm x_\ff(\bm x_\ii))}
{\partial(\bm x_\ii)}\,=\,\rho_\ii(\bm x_\ii)\,.
\label{crnt}
\qqq 
In other words, the Lagrangian map $\,\bm x_\ii\mapsto\bm x_\ff(\bm x_\ii)\,$ 
should transport the initial density $\,\rho_\ii\,$ into the 
final one $\,\rho_\ff$. 
\,Upon exchange of the order of integration, the minimization of functional 
(\ref{CF}) may be done in three steps: 
\begin{itemize}
\item First, we fix a smooth Lagrangian map  
\qq
\bm x_\ii\,\mapsto\,\bm x_\ff(\bm x_\ii)
\qqq
with a smooth inverse $\,\bm x_\ff\mapsto\bm x_\ii(\bm x_\ff)\,$ 
such that constraint (\ref{crnt}) holds. 
\item Second, for each $\,\bm x_\ii$, \,we minimize
\qq
\int_{0}^{t_\ff}\dot{\bm x}(t;\bm x_\ii)\cdot
M^{-1}\,\dot{\bm x}(t;\bm x_\ii)\,\,dt
\qqq
over the curves $\,[0,t_\ff]\ni t\,\mapsto\,\bm x(t;\bm x_\ii)\,$ 
starting from $\,\bm x_\ii\,$ and ending at $\,\bm x_\ff(\bm x_\ii)$.
\,Due to the positivity of matrix $\,M$, \,
the minimal curves are just the straight lines
\qq
[0,t_\ff]\ni t\ \longmapsto\ \bm x(t;\bm x_i)\,=\,\frac{t_\ff-t}{t_\ff}\,\bm x_\ii+
\frac{t}{t_\ff}\,\bm x_\ff(\bm x_\ii)
\label{minim1}
\qqq
with constant time-derivative 
$\,\dot{\bm x}(t;\bm x_\ii)=\bm x_\ff(\bm x_\ii)-\bm x_\ii$.  
\item Third, we minimize the ``quadratic cost functional''
\qq
\CK[\bm x_\ff(\cdot)]\,=\,\int(\bm x_\ff(x_\ii)-\bm x_\ii)\cdot{M}^{-1}
(\bm x_\ff(x_\ii)-\bm x_\ii)\,\,
\rho_\ii(\bm x_\ii)\,\,dx_\ii
\label{cost}
\qqq
over the maps $\,\bm x_\ii\mapsto\bm x_\ff(\bm x_\ii)\,$ satisfying
constraint (\ref{crnt}).
\end{itemize}

\noindent In principle, the above three-step minimization is over a broader 
class of maps $\,\bm x(t;\bm x_\ii)\,$ which might be non-invertible for 
fixed intermediate $\,t$, \,not representing the Lagrangian flow of any 
velocity field $\,\bm v(t,\bm x)$. \,As we shall see in the next 
section, however, the minimizer (\ref{minim1}) represents such a flow if 
$\,x_\ff(\bm x_\ii)\,$ minimizes the cost function (\ref{cost}) under 
constraint (\ref{crnt}).

\nsection{\bf Monge-Kantorovich mass transport and Burgers equation}
\label{sec:MK}

\noindent The minimization of the quadratic cost function (\ref{cost}) 
over invertible Lagrangian maps $\,\bm x_\ii\mapsto
\bm x_\ff(\bm x_\ii)\,$ satisfying constraint (\ref{crnt}) is the celebrated 
Monge-Kantorovich optimal mass transport problem \citep{Monge,Kantor} related
to the inviscid Burgers equation \citep{BB,BB1,V}. \,For reader's convenience,
we shall briefly recall that relation in the present section. 
\vskip 0.2cm

Observe that constraint (\ref{crnt}) may be rewritten in the equivalent form 
in terms of inverse Lagrangian maps as the identity
\qq
\rho_\ff(\bm x_\ff)\,=\,\rho_\ii(\bm x_\ii(\bm x_\ff))\,
\frac{\partial(\bm x_\ii(\bm x_\ff))}{\partial(\bm x_\ff)}\,.
\label{crnt1}
\qqq 
In the latter form, it implies for the infinitesimal variation
$\,\delta\bm x_\ii(\bm x_\ff)\,$ of the inverse Lagrangian map
the condition
\qq
\delta\bm x_\ii(\bm x_\ff)\cdot(\nabla_{\bm x_\ii}
\rho_\ii)(\bm x_\ii(\bm x_\ff))\,+\,\rho_\ii(\bm x_\ii(\bm x_\ff))
\,\frac{\partial x_\ff^a(\bm x_\ii)}{\partial x_\ii^b}
\,\frac{\partial\,\delta x^b_\ii(\bm x_\ff)}{\partial x_\ff^a}\ =\ 0\,,
\qqq
where the $2^{\rm nd}$ term comes from the variation of the Jacobian 
$\,\frac{\partial(\bm x_\ii(\bm x_\ff))}{\partial(\bm x_\ff)}$.
\,The last equation may be rewritten as a no-divergence requirement: 
\qq
\nabla_{\bm x_\ii}\cdot\big(\rho_\ii(\bm x_\ii)\,
\delta\bm x_\ii(\bm x_\ff(\bm x_\ii))\big)
 =\ 0\,.
\label{nodiv}
\qqq 
Changing variables in the expression (\ref{cost}) 
and using constraint (\ref{crnt1}), we may re-express the cost 
function in an equivalent form involving the final density:
\qq
\CK[\bm x_\ff(\cdot)]\,=\,\int(\bm x_\ff-\bm x_\ii(\bm x_\ff))\cdot{M}^{-1}
(\bm x_\ff-\bm x_\ii(\bm x_\ff)\,\,
\rho_\ff(\bm x_\ff)\,\,dx_\ff\,.
\label{cost1}
\qqq
The variation of the latter is
\qq
\delta\CK[\bm x_\ff(\cdot)]&=&2\int(\bm x_\ii(\bm x_\ff)-\bm x_\ff)\cdot
{M}^{-1}\delta\bm x_\ii(\bm x_\ff)\,\,\rho_\ff(\bm x_\ff)\,\,dx_\ff\cr
&=&2\int(\bm x_\ii-\bm x_\ff(\bm x_\ii))\cdot{M}^{-1}
\delta\bm x_\ii(\bm x_\ff(\bm x_\ii))\,\,\rho_\ii(\bm x_\ii)\,\,dx_\ii\,.
\label{variat}
\qqq
For the extremal maps $\,\bm x_\ii\mapsto\bm x_\ff(\bm x_\ii)$, \,variation
(\ref{variat}) has to vanish for all $\,\delta\bm x_\ii(\bm x_\ff(\bm x_\ii))\,$ 
satisfying (\ref{nodiv}). This occurs if and only if $\,{M}^{-1}(\bm x_\ii-
\bm x_\ff(\bm x_\ii))\,$ is a gradient, \,i.e. if there exists a function 
$\,F(\bm x_\ii)\,$ such that
\qq
\bm x_\ff(\bm x_\ii)\ =\ {M}\,\bm\nabla F(\bm x_\ii)\,.
\label{x0p0}
\qqq
Substituting this relation to expression 
(\ref{crnt}) for the constraint one infers that function $\,F\,$ 
solves the Monge-Amp\`ere equation
\qq
\rho_\ff\big({M}\,\bm\nabla F(\bm x_\ii)\big)\hspace{0.08cm}
\det\left({M}^{ac}\frac{\partial^2F}{\partial{x^b_\ii}\,\partial{x^c_\ii}}
(\bm x_\ii)\right)\ =\ \rho_\ii(\bm x_\ii)
\label{MA}
\qqq
and, in particular, that
\qq
\det\left({M}^{ac}\frac{\partial^2F}{\partial{x_\ii^b}\,\partial{x_\ii^c}}
(\bm x_\ii)\right)\ >\ 0
\label{Hp}
\qqq
(in the above relations, the mobility matrix $\,M\,$ may be absorbed by
the linear change of variables $\,\bm x\mapsto\bm x'=\sqrt{M}\,\bm x$).
\,The crucial input from the the theory of Monge-Kantorovich optimal mass 
transport is the result that the minimizer $\,\bm x_\ii\mapsto
\bm x_{\ff}(\bm x_\ii)\,$ of the cost function exists and is the unique 
extremum corresponding to a function $\,F\,$ which is convex \citep{GMcC,V}.  
\,Note that it follows then from Eq.\,(\ref{Hp}) that the Hessian matrix 
of $\,F\,$ is everywhere strictly positive. Now, interpolating between 
$\,\frac{1}{2}\,\bm x_\ii\cdot{M}^{-1}\bm x_\ii\,$ and function 
$\,F(\bm x_\ii)$, \,set
\qq
F_t(\bm x_\ii)\ =\ \frac{t_\ff-t}{2t_\ff}\,
\bm x_\ii\cdot{M}^{-1}\bm x_\ii\,+\,
\frac{t}{t_\ff}\,F(\bm x_\ii)
\qqq
for $\,0\leq t\leq t_\ff$. \,Hence 
\qq
{M}\,\bm\nabla F_t(\bm x_\ii)\,
=\,\frac{t_\ff-t}{t_\ff}\,\bm x_\ii\,+\,\frac{t}{t_\ff}\,
\bm x_\ff(\bm x_\ii)\,=\,\bm x(t;\bm x_\ii)\,,\quad
\label{xt}
\qqq
giving the linear interpolation between 
$\,\bm x_\ii\,$ and $\,\bm x_\ff(\bm x_\ii)$, \,just like in (\ref{minim1}).
\,Since
\qq
\frac{\partial\,x^{a}(t;\bm x_\ii)}{\partial x_\ii^b}(\bm x_\ii)\ 
=\ {M}^{ac}\frac{\partial^2F_t}{\partial{x_\ii^b}
\partial{x_\ii^c}}(\bm x_\ii)\ 
=\ \frac{t_\ff-t}{t_\ff}\,\delta^a_b
\,+\,\frac{t}{t_\ff}\,{M}^{ac}\frac{\partial^2F}
{\partial{x_\ii^b}\,\partial{x_\ii^c}}(\bm x_\ii)\,,
\qqq
it follows that matrix $\,\left(M^{-1}_{ab}
\frac{\partial x^b(t;\bm x_\ii)}{\partial x^c_\ii}\right)$,
\,equal to the Hessian matrix of $\,F_t$, \,is also everywhere positive
for the minimizer and even bounded below by the matrix 
$\,\frac{t_\ff-t}{t_\ff}M^{-1}$. 
\,This implies that the map $\,\bm x_\ii\mapsto\bm x(t;\bm x_\ii)\,$ 
is locally invertible and injective for all $\,t$.
\,The latter property is a consequence of the ``monotonicity''  
expressed by the inequalities
\qq
&&\big(\bm x^1_\ii-\bm x^0_\ii\big)\cdot M^{-1}\,\big(\bm x(t;\bm x^1_\ii)-
\bm x(t;\bm x^0_\ii)\big)\cr\cr
&&=\,\int_0^1\big(x^{1a}_\ii-x^{0a}_\ii\big)\,
\frac{\partial^2F_t}{\partial x^a_\ii\,\partial x^b_\ii}\big((1-s)
\bm x^0_\ii+s\bm x^1_\ii\big)\,\big(x^{1b}_\ii-x^{0b}_\ii\big)\,\,ds\cr\cr
&&\geq\ \frac{t_\ff-t}{t_\ff}\,(\bm x^1_\ii-\bm x^0_\ii)
\cdot M^{-1}(\bm x^1_\ii-\bm x^0_\ii)\ >\ 0
\qqq
which also imply that $\,\bm x(t,\bm x^1_\ii)\,$ has to sweep the whole
space when $\,(\bm x^1_\ii-\bm x^0_\ii)\cdot M^{-1}(\bm x^1_\ii-\bm x^0_\ii)\,$
increases from zero to infinity. \,Hence the global invertibility of the 
maps $\,\bm x_\ii\mapsto\bm x(t;\bm x_\ii)$.
\,It then makes sense to define a function $\,\Psi(t,\bm x)\,$ by the 
relation
\qq
\Psi(t,\bm x)\ =\ \frac{1}{t}\Big[\frac{1}{2}\hspace{0.03cm}
\bm x\cdot M^{-1}\bm x\,-\,\bm x\cdot M^{-1}\bm x_\ii\,+\,F_t(\bm x_\ii)
\Big]_{{\bm x(t;\bm x_\ii)=\bm x}}.
\label{psit}
\qqq
Note that the derivative over $\,\bm x_\ii\,$ of the term
$\,[\,\cdots\,]\,$ on the right hand side vanishes 
for $\,{{\bm x(t;\bm x_\ii)=\bm x}}\,$ due to Eq.\,(\ref{xt}). 
\,It follows that
\qq
\partial_t\Psi(t,\bm x)
=-\,\frac{1}{2t^2}\,(\bm x-\bm x_\ii)\cdot
{M}^{-1}(\bm x-\bm x_\ii)\,,\qquad
{\bm\nabla}\Psi(t,\bm x)=\frac{1}{t}\,
{M}^{-1}\big(\bm x-\bm x_\ii\big)
\label{nn}
\qqq
for $\,{\bm x(t;\bm x_\ii)=\bm x}$.
\,Comparing the last two equations, we infer that function 
$\,\Psi(t,\bm x)\,$ satisfies the non-linear evolution equation
\qq
\partial_t\Psi\,+\,\frac{1}{2}\,({\bm\nabla}\Psi)\cdot M\,({\bm\nabla}
\Psi)\,=\,0
\label{pBurg}
\qqq
that implies the inviscid  Burgers equation (the Euler equation without
pressure)
\qq
\partial_t\bm v\,+\,(\bm v\cdot{\bm\nabla})\,\bm v\,=\,0
\label{Burg}
\qqq 
for the velocity field 
\qq
\bm v(t,\bm x)\,=\,M\,{\bm\nabla}\Psi(t,\bm x)\,.
\label{v*}
\qqq
Eqs.\,(\ref{xt}) and (\ref{nn}) entail that 
\qq
\dot{\bm x}(t;\bm x_\ii)\,=\,\frac{1}{t_\ff}\,
\big(\bm x_\ff(\bm x_\ii)-\bm x_\ii\big)\,=\,\frac{1}{t}\,
\big(\bm x(t;\bm x_\ii)-\bm x_\ii)\,=\,
{M}\,({\bm\nabla}\Psi)(t,\bm x(t;\bm x_\ii))\,=\,
\bm v(t,\bm x(s;\bm x_\ii))\quad
\label{Lfv}
\qqq
so that the interpolating maps $\,\bm x(t;\bm x)\,$ provide
the Lagrangian flow of the Burgers velocity field $\,\bm v\,$ and that
the latter is conserved along that flow. This is a general fact:
Lagrangian trajectories of a velocity field solving the inviscid Burgers 
equation have constant velocities.  
\vskip 0.3cm

Let us define the intermediate densities $\,\rho(t,\bm x)\,$ that 
interpolate over the time interval $\,[0,t_\ff]\,$ between $\,\rho_\ii\,$ 
and $\,\rho_\ff\,$ by Eq.\,(\ref{sFP}) so that they evolve according 
to the advection equation (\ref{rhoadv}) in the Burgers velocity field 
$\,\bm v\,$ of Eq.\,(\ref{v*}). \,It is the assumption that the initial 
and final densities are smooth that assures that such velocities do 
not involve shocks on the time interval $\,[0,t_\ff]$. 
\vskip 0.3cm

Summarizing the above discussion, we infer that the Burgers velocity field 
$\,v(t,\bm x)\,$ of Eq.\,(\ref{v*}), together with the densities
$\,\rho(t,\bm x)\,$ of Eq.\,(\ref{sFP}), \,minimize functional 
$\,\CA[\bm v,\rho_\ii]\,$ of Eq.\,(\ref{CS}) over the space of velocities 
$\,\bm v(t,\bm x)\,$ and 
densities $\,\rho(t,\bm x)\,$ that evolve for $\,0\leq t\leq t_\ff\,$ 
by the advection equation (\ref{rhoadv}) interpolating between 
$\,\rho_\ii\,$ and $\,\rho_\ff$. \,The minimal value of functional 
$\,\CA[\bm v,\rho_\ii]\,$ under the above constraint is
\qq
\CA_{min}\,=\,\frac{1}{t_\ff T}\hspace{0.03cm}\,\CK_{min}\,,
\label{CSm}
\qqq 
where $\,\CK_{min}\,$ is the value of the quadratic cost
function (\ref{cost}) on the minimizer $\,\bm x_\ii\mapsto\bm x_\ff(x_\ii)$.
\,These are the main results of \citep{BB,BB1}, \,see also Chapter 8 of \citep{V} 
for more details. That $\,\CA_{min}\,$ had to be inversely proportional to 
the length of the time interval could have been inferred directly by 
rescaling of time in functional (\ref{CS}) \citep{SS1}.
\vskip 0.3cm

Below, we shall use the following factorization property
of the optimal mass transport problem with the cost function (\ref{cost})
holding if the mobility matrix has the block form
\qq
M\,=\,\Big(\begin{matrix}{M^1&0\cr0&M^2}\end{matrix}\Big).
\label{block}
\qqq
If, with respect to the corresponding decomposition of the $\,d$-dimensional
space, both initial and final densities have the product form:
\qq
\rho_\ii(\bm x)\,=\,\rho_\ii^1(\bm x^1)\,\rho^2_\ii(\bm x^2)\,,\qquad
\rho_\ff(\bm x)\,=\,\rho_\ff^1(\bm x^1)\,\rho^2_\ff(\bm x^2)
\qqq
for $\,\bm x=(\bm x^1,\bm x^2)$, \,then the Lagrangian map minimizing 
cost function (\ref{cost}) also factorizes into the product of minimizers 
of the lower dimensional problems:
\qq
\bm x_\ff(\bm x_\ii)\,=\,M\,\bm\nabla F(\bm x_\ii)\ =\ 
\big(\bm x^{1}_\ff(\bm x_\ii^1),\,\bm x^{2}_\ff(\bm x_\ii^1)\big)\,=\,
\big(M^1\bm\nabla F^{1}(\bm x_\ii^1),\,M^2\bm\nabla F^{2}
(\bm x_\ii^2)\big)    
\label{fact}
\qqq
and the minimal cost is the sum of the lower-dimensional ones.
This follows from the uniqueness of the minimizer and its characterization
in terms of the gradient of a convex function. The corresponding
Burgers potential $\,\Psi(t,\bm x)\,$ is then the sum, and the interpolating 
density $\,\rho(t,\bm x)\,$ the product, of the ones obtained from 
the lower dimensional minimizers.

\nsection{Second Law of Stochastic Thermodynamics at short times}
\label{sec:R2ndLaw}

\noindent Let us denote by $\,R(t,\bm x)\,$ the dynamic potential
related by Eq.\,(\ref{Rt}) to the optimal densities $\,\rho(t,\bm x)$.
\,Set 
\qq
U(t,\bm x)\,=\,R(t,\bm x)-\Psi(t,\bm x)\,,
\label{optp}
\qqq
where $\,\Psi\,$ is the Burgers potential (\ref{psit}).
\,Eq.\,(\ref{v*}) for the optimal velocity may be rewritten as
\qq
\bm v\,=\,M\,{\bm\nabla}(R-U)\,.
\label{v*U}
\qqq
and the advection equation (\ref{rhoadv}) for $\,\rho\,$
becomes
\qq
\partial_t\rho\ =\ -\,M\,{\bm\nabla}(\rho{\bm\nabla}(R-U))
\ =\ L_t^\dagger\rho\,,
\qqq
where $\,L_t\,$ is the time-dependent generator (\ref{Lt})
for the Langevin process with control $\,U$. \,We infer that 
the optimal $\,\rho\,$ describes the instantaneous probability densities   
of such a process with initial values distributed with density 
$\,\rho_\ii$ and that the optimal $\,\bm v\,$ is its mean local velocity. 
\,It follows then from relation (\ref{DStot}) that control 
$\,U\,$ provides the optimal protocol on the time interval $\,[0,t_\ff]\,$ 
that evolves the initial state $\,\rho_\ii\,$ 
to the final state $\,\rho_\ff\,$ under the Langevin dynamics (\ref{Lan})
with the minimal total entropy production equal to $\,\CA_{min}\,$ of 
\,Eq.\,(\ref{CSm}). \,We obtain this way a refinement for finite time 
intervals of the Second Law (\ref{2ndLaw}) of Stochastic Thermodynamics: 
\vskip 0.5cm

\noindent{\bf Theorem.} \ \,\parbox[t]{13.6cm}{For the Langevin dynamics 
(\ref{Lan}) on the time interval $\,[0,t_\ff]\,$ that evolves between
states $\,\rho_\ii\,$ and $\,\rho_\ff$,
\qq
\Delta S_{tot}\ \geq\ 
\frac{1}{t_\ff T}\hspace{0.08cm}\CK_{min}\,,
\label{R2ndLaw}
\qqq
with the inequality saturated by the optimal evolution with the 
time dependent potential $\,U(t,\bm x)\,$ constructed above.}  
\vskip 0.5cm

\noindent Here, as in relation (\ref{2ndLaw}), 
$\,\Delta S_{tot}=\Delta S_{sys}+
\Delta S_{env}\,$ denotes the total entropy change, composed of the
change of entropy of the system $\,\Delta S_{sys}\,$  
and the change of entropy of the thermal environment
$\,\Delta S_{env}=\frac{1}{T}\big\langle\,Q\,\big\rangle\,$
during the process. The theorem states that the total change of entropy 
during Langevin evolution (\ref{Lan}) is not smaller than the minimal 
quadratic cost function (involving the mobility matrix $\,{M}$) \,for 
the transport of initial probability distribution to the final one, 
divided by the product of time length $\,t_\ff\,$ of the process 
by temperature $\,T\,$ of the environment. Since the cost function
is strictly positive whenever the initial and final probability 
distributions are different, it follows that the shorter the time 
length of the process and the smaller temperature, the bigger
minimal total entropy production. The latter may approach zero
only for (adiabatically slow) processes taking very long time.
Inequality (\ref{R2ndLaw}) provides then a quantitative refinement 
of the Second Law of Stochastic Thermodynamics (\ref{2ndLaw}) 
for processes whose time span does not exceed $\,t_\ff$. 
\,In order to determine the optimal protocol $\,U(t,\bm x)\,$ 
of Eq.\,(\ref{optp}), \,one has to find subsequently: 
\begin{enumerate}
\item the minimizer $\,\bm x_\ii\mapsto\bm x_\ff(\bm x_\ii)=
M\,\bm\nabla F(\bm x_\ii)\,$ of the cost function of Eq.\,(\ref{cost})
under the constraint (\ref{crnt}) such that $\,\CK_{min}=\CK[\bm x_\ff(\cdot)]$;
\item the solution $\,\Psi\,$ given by Eq.\,(\ref{psit}) of the
Burgers equation (\ref{pBurg}) for potentials;
\item the solution $\,\rho\,$ given by Eq.\,(\ref{sFP})
of the advection equation (\ref{rhoadv}) in the Burgers velocity field
$\,\bm v=M\,{\bm\nabla}\Psi$.
\end{enumerate} 
\vskip 0.2cm

The refined Second Law (\ref{R2ndLaw}) may be rewritten as a refinement 
of the lower bound (\ref{bdforQ}) for the heat release in processes
with fixed initial and final densities that takes the form
\qq
\big\langle\,Q\,\big\rangle\ \geq\ \frac{1}{t_\ff}\hspace{0.1cm}\CK_{min}
\,-\,T\,\Delta S_{sys}\,.
\label{RbdforQ}
\qqq
and is saturated for the same optimal protocol that the inequality 
(\ref{R2ndLaw}). \,If one admits initial and final jumps of control $\,U_t$,
\,as discussed in Sec.\,\ref{sec:secondLaw}, \,then the problem
considered in \citep{SS,AMMMG} of minimization of average work 
(\ref{avW1}) for fixed initial control $\,U_\ii$, \,initial 
density $\,\rho_\ii$, \,and final control $\,U_\ff$, 
\,but for arbitrary final density $\,\rho_\ff$, \,is very closely 
related to the problem of minimizing the heat release. Indeed, we may 
first minimize $\,\langle\,Q\,\rangle\,$ for fixed $\,\rho_\ii\,$ and 
$\,\rho_\ff\,$ and then minimize the right hand side of Eq.\,(\ref{avW1}) 
over $\,\rho_\ff$. \,This gives the inequality
\qq
\big\langle\,W\,\big\rangle\ \geq\ \mathop{{\rm min}}\limits_{\rho_\ff}
\Big[\frac{1}{t_\ff}\hspace{0.1cm}\CK_{min}\,
+\,\,\int(U_\ff- R_\ff)(\bm x)\,\,\rho_\ff(\bm x)\,\,dx\Big]\,-\,
\int(U_\ii- R_\ii)(\bm x)\,\,\rho_\ii(\bm x)\,\,dx\,,
\label{minW}
\qqq
where, as before, $\,\CK_{min}\,$ denotes the minimal value of the cost 
function (\ref{cost}) for the transport of $\,\rho_\ii\,$ to $\,\rho_\ff$.
The above bound is saturated for the protocol $\,U(t,\bm x)\,$ that 
minimizes the average heat release for the fixed final density $\,\rho_\ff\,$
corresponding to the minimizer of the expression in the square brackets 
on the right hand side. 
\vskip 0.5cm

\noindent{\bf Example.} \ \ For the Gaussian example (discussed in 
\citep{AMMMG}, see also \citep{SS}) with $\,M^{ab}=\mu\,\delta^{ab}$,
\,take
\qq
\rho_\ii(\bm x)\ =\ \frac{1}{(2\pi\sigma_\ii^2)^{d/2}}\,
\exp\Big[-\frac{(\bm x-\bm\alpha_\ii)^2}{2\sigma_\ii^2}\Big]\,,
\qquad\rho_\ff(\bm x)\ =\ \frac{1}{(2\pi\sigma_\ff^2)^{d/2}}\,
\exp\Big[-\frac{(\bm x-\bm\alpha_\ff)^2}{2\sigma_\ff^2}\Big]\,.
\label{rhoirhofg}
\qqq
corresponding to the system entropy change 
$\,\Delta S_{sys}=d\,k_B\ln(\sigma_\ff/\sigma_\ii)$. \,The optimal Lagrangian map 
in this case is linear:
\qq
\bm x_\ff(\bm x_\ii)&=&\frac{\sigma_\ff}{\sigma_\ii}(\bm x_\ii
-\bm\alpha_\ii)+\bm\alpha_\ff\,=\,\mu\,\bm\nabla F(\bm x_\ii)\,,
\qqq
with quadratic function
\qq
F(\bm x_\ii)\,=\,\frac{\sigma_\ff}{2\hspace{0.03cm}\sigma_\ii
\hspace{0.01cm}\mu}\big(\bm x_\ii
-\bm\alpha_\ii+\frac{\sigma_\ii}{\sigma_\ff}\bm\alpha_\ff\big)^2\,.
\qqq
Up to a constant, \,the solution of the Burgers equation (\ref{pBurg}) 
for potential is
\qq
\Psi(t,\bm x)\ =\ \frac{1}{2\,\mu\,t_\ff\,\sigma_t}
\Big((\sigma_\ff-\sigma_\ii)\,\bm x^2
\,+\,(\sigma_\ii\bm\alpha_\ff-\sigma_\ff\bm\alpha_\ii)\cdot
(2\hspace{0.02cm}\bm x-\bm\alpha_t)\Big),\quad
\label{psitt}
\qqq
\vskip -0.3cm
\noindent where
\vskip -0.5cm
\qq
\sigma_t=\frac{{t_\ff-t}}{{t_\ff}}\,\sigma_\ii+
\frac{{t}}{{t_\ff}}\,\sigma_\ff\,,\qquad\bm\alpha_t\ =\ 
\frac{{t_\ff-t}}{{t_\ff}}\,\bm\alpha_\ii+
\frac{{t}}{{t_\ff}}\,\bm\alpha_\ff
\label{stmt}
\qqq
interpolate linearly between the limiting values, 
\,and the intermediate probability densities become
\qq
\rho(t,\bm x)\ =\ \frac{1}{(2\pi\sigma_t^2)^{d/2}}\,
\exp\Big[-\frac{(\bm x-\bm\alpha_t)^2}{2\sigma_t^2}\Big],
\label{rhottt}
\qqq
Up to a time-dependent constant, \,the optimal control has the form
\qq
U(t,\bm x)\ =\ \frac{k_BT\mu\,t_\ff-\sigma_t
(\sigma_\ff-\sigma_\ii)}{{2\sigma_t^2\mu\,t_\ff}}
\Big(\bm x\,-\,\frac{k_BT\mu\,t_\ff\,\bm\alpha_t
+\sigma_t(\sigma_\ii\bm\alpha_\ff-\sigma_\ff\bm\alpha_\ii)}
{k_BT\mu\,t_\ff-\sigma_t(\sigma_\ff-\sigma_\ii)}\Big)^2.
\label{opcon}
\qqq
The minimal quadratic cost function is
\qq
\CK_{min}\ =\ \frac{(\sigma_\ff-\sigma_\ii)^2d+
(\bm\alpha_\ff-\bm\alpha_\ii)^2}{\mu}\,.
 \qqq
It determines the minimal total entropy production and minimal
average heat release saturating inequalities 
(\ref{R2ndLaw}) and (\ref{RbdforQ}). \,If, instead of the 
final density $\,\rho_\ff$, \,we fix the final control 
$\,U_\ff(\bm x)=\frac{(\bm x-\bm\alpha'_\ff)^2}{2\sigma'^2_\ff}$,
\,admitting its jump at $\,t=t_\ff\,$ then the minimum of the
average work (\ref{avW1}) is given by the right hand side of 
inequality (\ref{minW}), where the minimum over $\,\rho_\ff\,$
is attained on the Gaussian distribution $\,\rho_\ff\,$  
of Eqs.\,(\ref{rhoirhofg}) with
\qq
\sigma_\ff=\frac{2\sigma'^2_\ff\sigma_\ii
\,+\,\sqrt{(2\sigma'^2_\ff\sigma_\ii)^2\,+\,4k_BT\mu\,t_\ff\,
\sigma'^2_\ff(2\sigma'^2_\ff+k_BT\mu\,t_\ff)}}
{2(2\sigma'^2_\ff+k_BT\mu\,t_\ff)}\,,\quad
\bm\alpha_\ff=\frac{2\sigma'^2_\ff\,\bm\alpha_\ii
+k_BT\mu\,t_\ff\,\bm\alpha'_\ff}{2\sigma'^2_\ff
+k_BT\mu\,t_\ff}\,.\qquad
\qqq

\nsection{Finite time refinement of Landauer principle}
\label{sec:RLandauer}

\noindent The finite time refinement (\ref{RbdforQ}) of the lower
bound (\ref{bdforQ}) for the average heat release implies immediately 
a refinement of the Landauer bound for the average heat dissipated 
during the memory erasure of one bit of information in Langevin processes 
for which such erasure is related to the change $\,\Delta S_{sys}
=\,-\hspace{0.02cm}(\ln{\hspace{-0.04cm}2})\hspace{0.03cm}k_B\,$ 
of the the entropy of the system, see Sec.\,\ref{sec:secondLaw}. 
The improved bound takes the form 
\qq
\big\langle\,Q\,\big\rangle\ \geq\ \frac{1}{t_\ff}\,\CK_{min}\,+\,
(\ln{\hspace{-0.04cm}2})\,k_BT\,,
\label{RLand}
\qqq
where $\,\CK_{min}\,$ is the minimal value of the cost function (\ref{cost}).
\,In \citep{DLutz}, the distribution of the released heat (and work) was studied
numerically for a particular memory erasure overdamped one-dimensional
Langevin dynamics. It was checked that the mean heat release 
$\,\big\langle\,Q\,\big\rangle\,$ satisfied the Landauer bound, but that, 
with small but sizable probability, the fluctuating values of $\,Q\,$ 
may violate the bound. In \citep{Berut}, heat release was studied 
in an experimental realization of a similar system undergoing a memory erasure 
dynamics. In the experiment, a silica ball with diameter of $\,2\,\mu m\,$ 
suspended in a flat horizontal cell with ultra pure water at room 
temperature was manipulated by laser tweezers in order to displace 
the particle localized initially in a double trap to a fixed one 
of two traps. It was noticed in \citep{Berut} (in Fig.\,13) that, for 
a specific dynamical protocol, the difference between the mean heat release 
and the Landauer lower bound decreased with the time length $\,t_\ff\,$ of 
the erasure process (the decrease seemed inversely proportional to $\,t_\ff$). 
\,In order to see how the optimal protocol for which the upper bound 
in (\ref{RLand}) is saturated looks like in the experimental situation, we 
considered a 1-dimensional stochastic evolution (\ref{Lan}) with mobility 
$\,\mu=\frac{0.213877}{k_BT}\,\frac{\mu m^2}{s}\,$ and the limiting 
distributions
\qq
\rho_\ii(x)\hspace{0.03cm}&\,=\,&\frac{_1}{^{Z_\ii}}\,
\exp{\big[-\frac{_A}{^{k_BT}}(x^2-\alpha^2)^2\big]}
\ \equiv\ \exp{\big[-\frac{_1}{^{k_BT}}\,R_\ii(x)\big]}\,,\label{rhoii}\\
\rho_\ff(x)&\,=\,&\frac{_1}{^{Z_\ff}}
\,\exp{\big[-\frac{_A}{^{k_BT}}(x-\alpha)^2((x-\alpha)^2+3\alpha
(x-\alpha)+4\alpha^2)\big]}
\ \equiv\ \exp{\big[-\frac{_1}{^{k_BT}}\,R_\ff(x)\big]},\quad\label{rhoff}
\qqq
for $\,A=112\,k_BT\,\mu m^{-4}$, $\,\alpha=0.5\,\mu m$, and $\,x\,$ expressed
in $\,\mu m$'s, \,see Figs.\,1 and 2. The entropy difference between 
$\,\rho_\ii\,$ and $\,\rho_\ff\,$ is
\qq
\Delta S\,\approx\,-\hspace{0.03cm}0.7431204\,k_B 
\label{entrd}
\qqq
which is equal to $\,-\hspace{0.03cm}(\ln{\hspace{-0.04cm}2})\hspace{0.04cm}
k_B\,$ within $\,7.3\%$.

\begin{figure}[h]
\begin{center}
\leavevmode
\hspace{0.3cm}
{%
      \begin{minipage}{0.4\textwidth}\hspace*{-0.4cm}
        \includegraphics[width=7cm,height=4.5cm,angle=0]{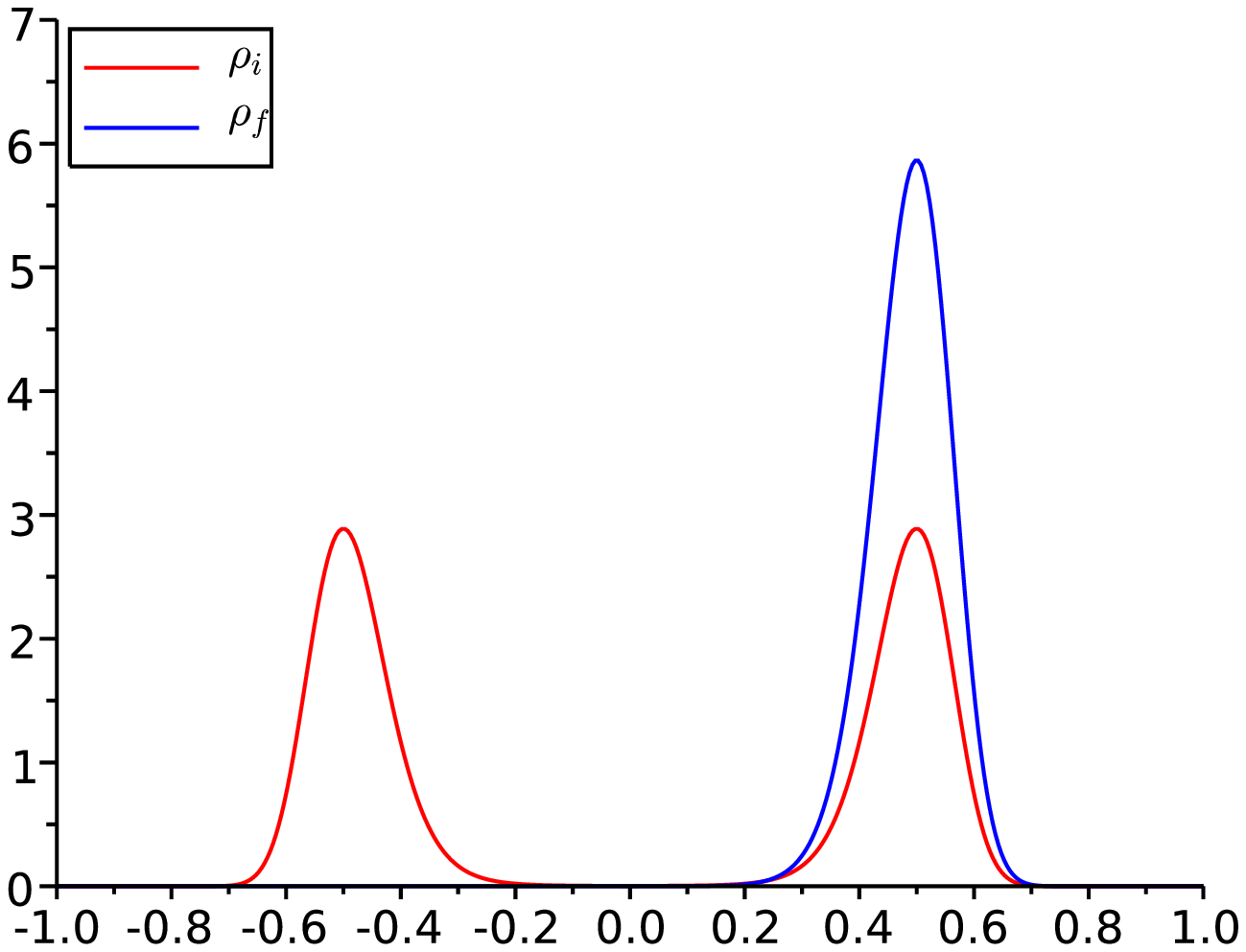}\\
        \vspace{-0.73cm} \strut
        \caption{\ $\rho_\ii$ {and} $\rho_\ff$}   
        \end{minipage}}
    \hspace*{0.5cm}
{%
      \begin{minipage}{0.4\textwidth}\vspace{0.17cm}\hspace*{-0.3cm}
        \includegraphics[width=7cm,height=4.4cm]{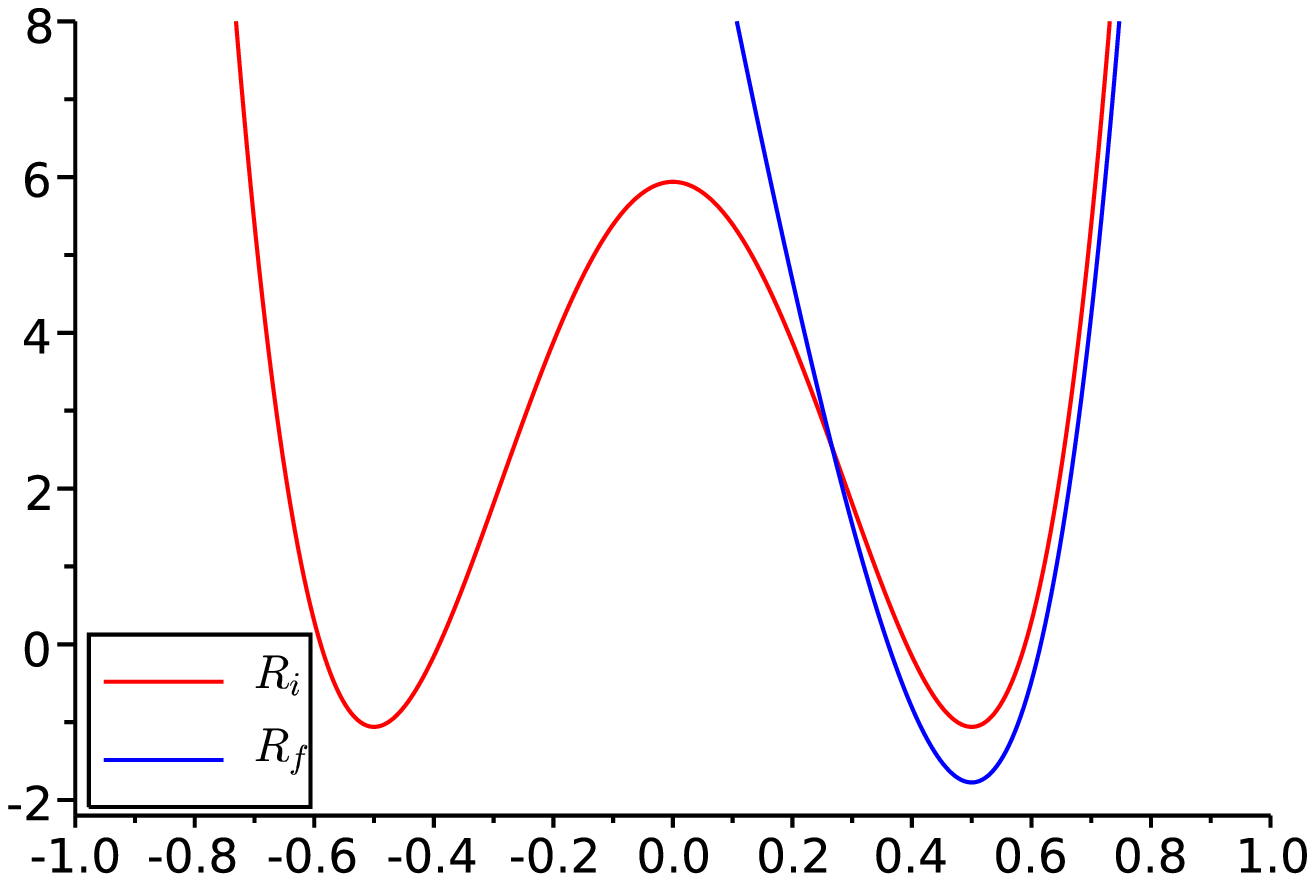}\\
        \vspace{-0.7cm} \strut
        \caption{\ $R_\ii$ and $R_\ff$}
        \end{minipage}}
    \hspace*{7pt}
\end{center}
\end{figure}
\vskip 0.1cm

The experimental situation is close to a two-dimensional one in 
the horizontal plane, where the above initial and final densities in the 
direction of $\,x$-axis are multiplied by the same density
\qq
\rho(y)\ =\ \frac{_1}{^Z}\,\exp{\big[-\frac{_A}{^{k_BT}}y^2(y^2+4\alpha^2)\big]}
\label{ydensity}
\qqq
in the direction of $\,y$-axis, leading to the same entropy difference
(\ref{entrd}). In such a two-dimensional situation, however, the 
Lagrangian map for the optimal mass transport factorizes into the map
$\,x_\ii\mapsto x_\ff(x_\ii)\,$ giving the optimal transport of 
$\,\rho_\ii(x)\,$ to $\,\rho_\ff(x)\,$ times the identity map in the
$\,y$-direction, see the end of Sec.\,\ref{sec:MK}. The minimal cost 
for the 2-dimensional problem coincides then with the one for the map 
$\,x_\ff(x_\ii)$. \,The corresponding two-dimensional optimal control is 
the sum of the optimal control $\,U(t,x)\,$ for the one-dimensional 
problem in the $\,x$-direction and of the static potential 
$\,U(y)=A\,y^2(y^2+4\alpha^2y^2)$. \,Consequently, the two-dimensional problem 
reduces to the one-dimensional one in the direction of the $\,x$-axis. 
Similarly, the strong confining potential in the vertical $\,z$-direction 
may be ignored as long as it is $\,x$- and $\,y$-independent.
\vskip 0.2cm

We employ three methods to find the optimal Lagrangian map 
$\,x_\ii\mapsto x_\ff(x_\ii)\,$  that transports 
$\,\rho_\ii\,$ to $\,\rho_\ff\,$ and minimizes the quadratic 
cost. \,First, the unique positively oriented map 
$\,x_\ii\mapsto x_\ff(x_\ii)\,$ that transports $\,\rho_\ii\,$ to 
$\,\rho_\ff\,$  may be found from the relation
\qq
\int_{-\infty}^{\,x_{^\ff}(x_\ii)}\hspace{-0.2cm}\rho_\ff(x)\,dx\ 
=\ \int_{-\infty}^{\,x_\ii}\hspace{-0.1cm}\rho_\ii(x)
\,dx\,.
\label{baseq}
\qqq
According to the general theory exposed in Sec.\,\ref{sec:MK}, \,it has 
to minimize the quadratic cost since it is a gradient of a convex function. 
We solve Eq.\,(\ref{baseq}) for $\,x_\ff(x_\ii)\,$ numerically in Scilab, 
using the {\,\it fsolve\,} procedure, for discrete values of $\,x_\ii\,$ spaced 
by 5\,nm lying in the interval $\,-0.7725\,\mu m\leq x_\ii\leq 0.6925\,\mu m$.
\,This method did not give access to the remaining values of $\,x_\ii$, 
\,the numbers involved exceeding there the program capacity. \,Instead, 
\,for $\,x_\ii<-0.7725\,\mu m$, \,the Lagrangian map $\,x_\ii\mapsto 
x_\ff(x_\ii)\,$ was calculated by expanding $\,(x_\ff(x_\ii)-x_\ii)\,$ 
in powers of $\,(x_\ii-\alpha)^{-1}\,$ up to order 11. The coefficients of 
the expansion were found from the derivative equation
\qq
\frac{{dx_\ff(x_\ii)}}{{dx_\ii}}\ =\ \frac{{\rho_\ii(x_\ii)}}
{{\rho_\ff(x_\ff(x_\ii))}}\,.
\qqq
Similarly, for $\,x>0.6925\,\mu m$, \,the map $\,x_\ii\mapsto 
x_\ff(x_\ii)\,$ was computed by expanding $\,(x_\ff(x_\ii)-x_\ii)\,$ in 
powers of $\,(x_\ii+\alpha)^{-1}$. 
\,Finally, in order to check the above results, in particular around the 
boundary points of the $\,x_\ii$-intervals, where they become less reliable,
we performed numerical search for the solution of the corresponding optimal 
assignment problem, usually employed in numerical optimization of 
higher-dimensional mass transport \citep{BFHLMMS}. 
The task is to find the permutation $\,\pi\,$ of length $\,N\,$ 
that induces a bijective map $\,\bm q_n\mapsto\bm x_{\pi(n)}\,$ between 
$\,N\,$ points (``particles'') $\,\bm q_n\,$ distributed with density 
$\,\rho_\ii\,$ and $\,N\,$ points $\,\bm x_n\,$ distributed with density 
$\,\rho_\ff$, minimizing the discretized quadratic cost
\qq
K^{disc}\ =\ \sum\limits_{n=1}^N (\bm x_{\pi(n)}-\bm q_n)
\cdot M^{-1}(\bm x_{\pi(n)}-\bm q_n)
\qqq

\begin{figure}[t]
\leavevmode
\begin{center}
\vskip -0.3cm
\includegraphics*[width=8.5cm,height=6.8cm,angle=0]{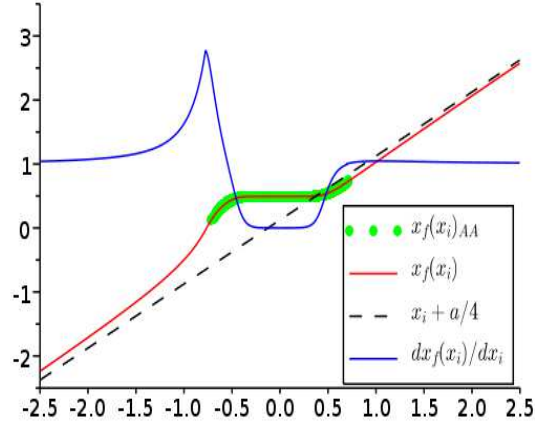}\\
        \vskip -0.3cm
\caption{\ Lagrangian map $\,x_\ff(x_\ii)$, \,its asymptotes and 
                   its derivative}
\end{center}
\end{figure}
\vskip 0.2cm

\begin{figure}[ht]
\leavevmode
\begin{center}
\vskip -0.3cm
{%
      \begin{minipage}{0.4\textwidth}\hspace*{-0.9cm}
        \includegraphics[width=7.8cm,height=5.7cm,angle=0]{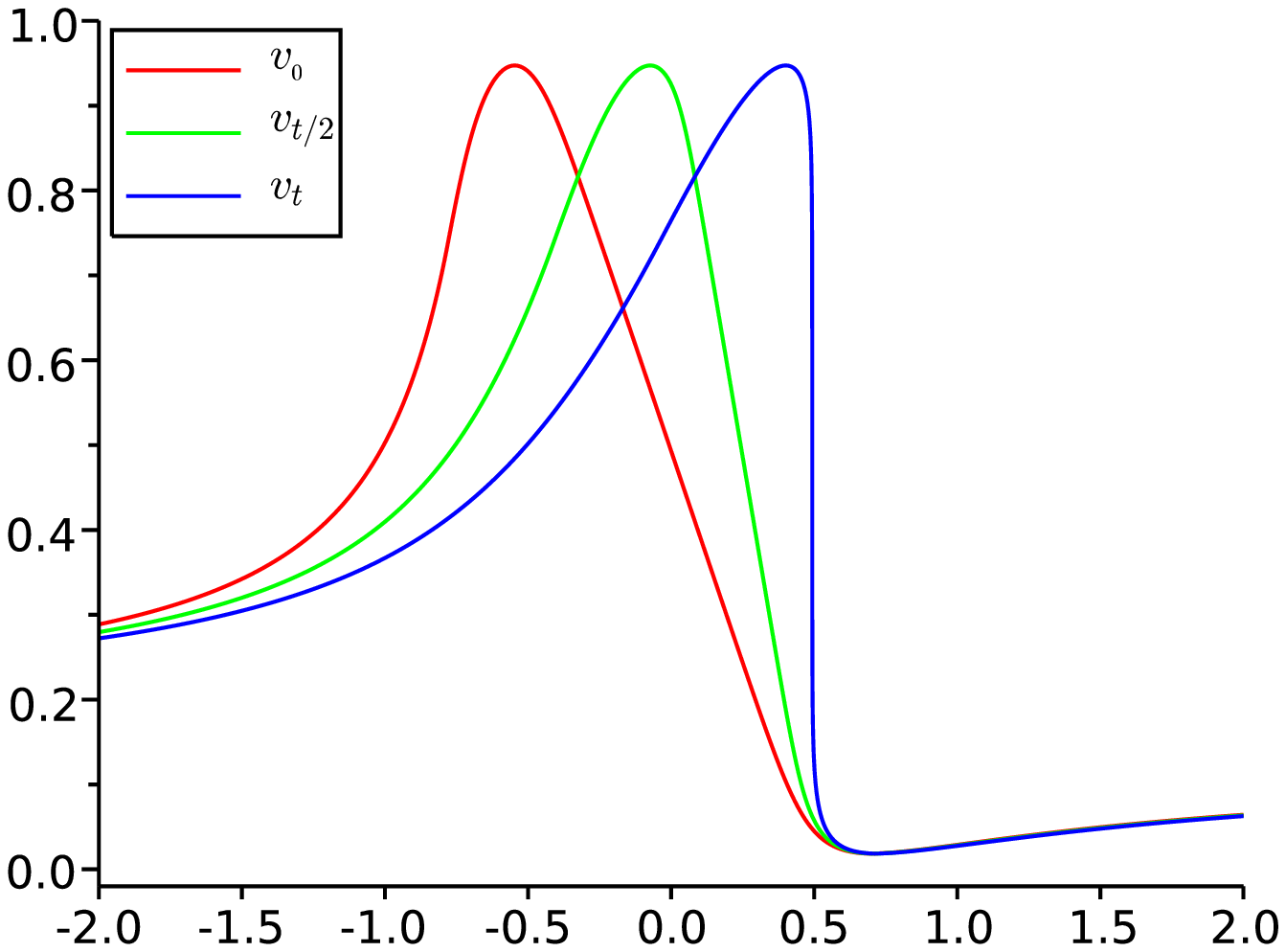}\\
        \vspace{-0.73cm} \strut
        \caption{Burgers velocity at initial,\\
                 \hspace*{1.2cm}half-time and final times,\\ 
                 \hspace*{1.2cm}time window $\,t_\ff=1s$}   
        \end{minipage}}
    \hspace*{0.8cm}
{%
      \begin{minipage}{0.4\textwidth}\vspace{0.17cm}\hspace*{-0.6cm}
        \includegraphics[width=7.8cm,height=5.7cm]{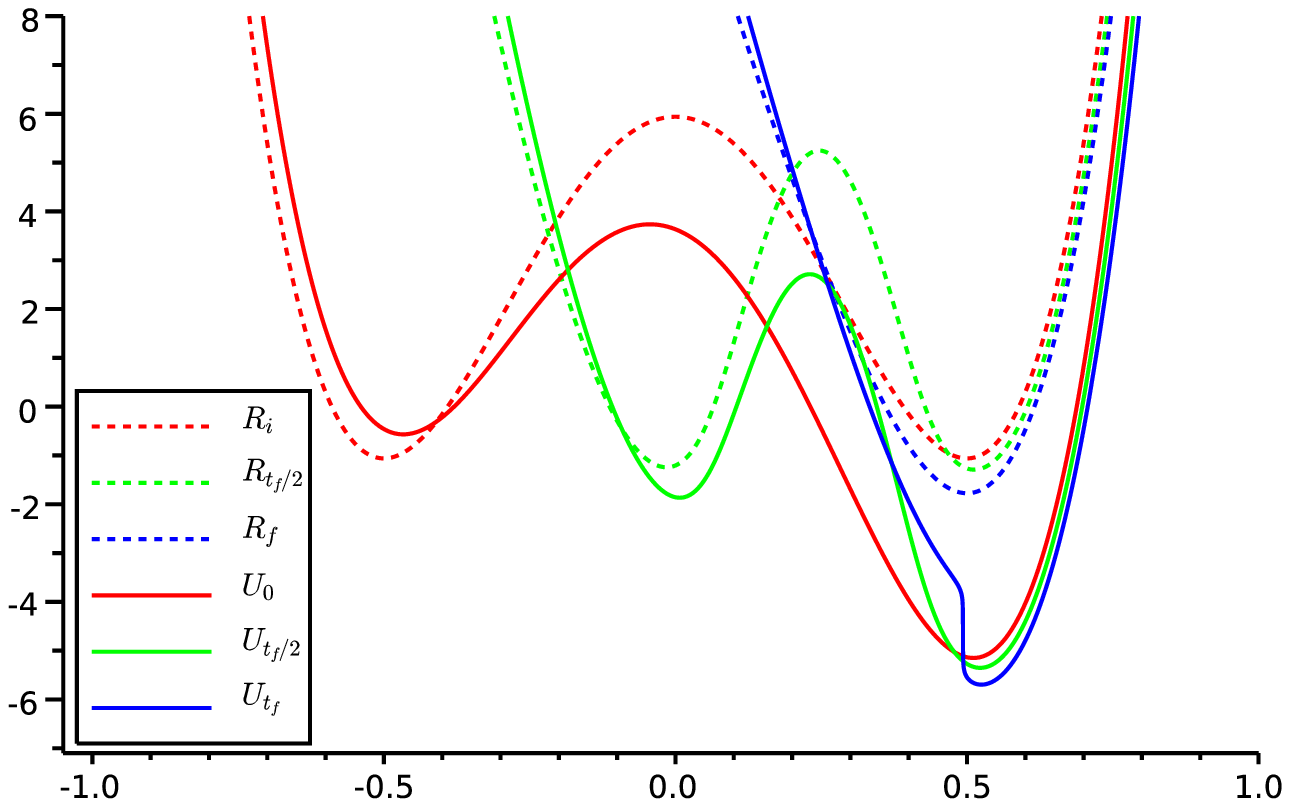}\\
        \vspace{-0.7cm} \strut
        \caption{\ initial, half-time and final potentials\\
                 \hspace*{1.2cm}initial, half-time and final controls,\\
                 \hspace*{1.2cm}time window $\,t_\ff=1s$}   
        \end{minipage}}
    \hspace*{7pt}
\vskip 0.3cm
{%
      \begin{minipage}{0.4\textwidth}\hspace*{-0.9cm}
        \includegraphics[width=7.8cm,height=5.7cm,angle=0]{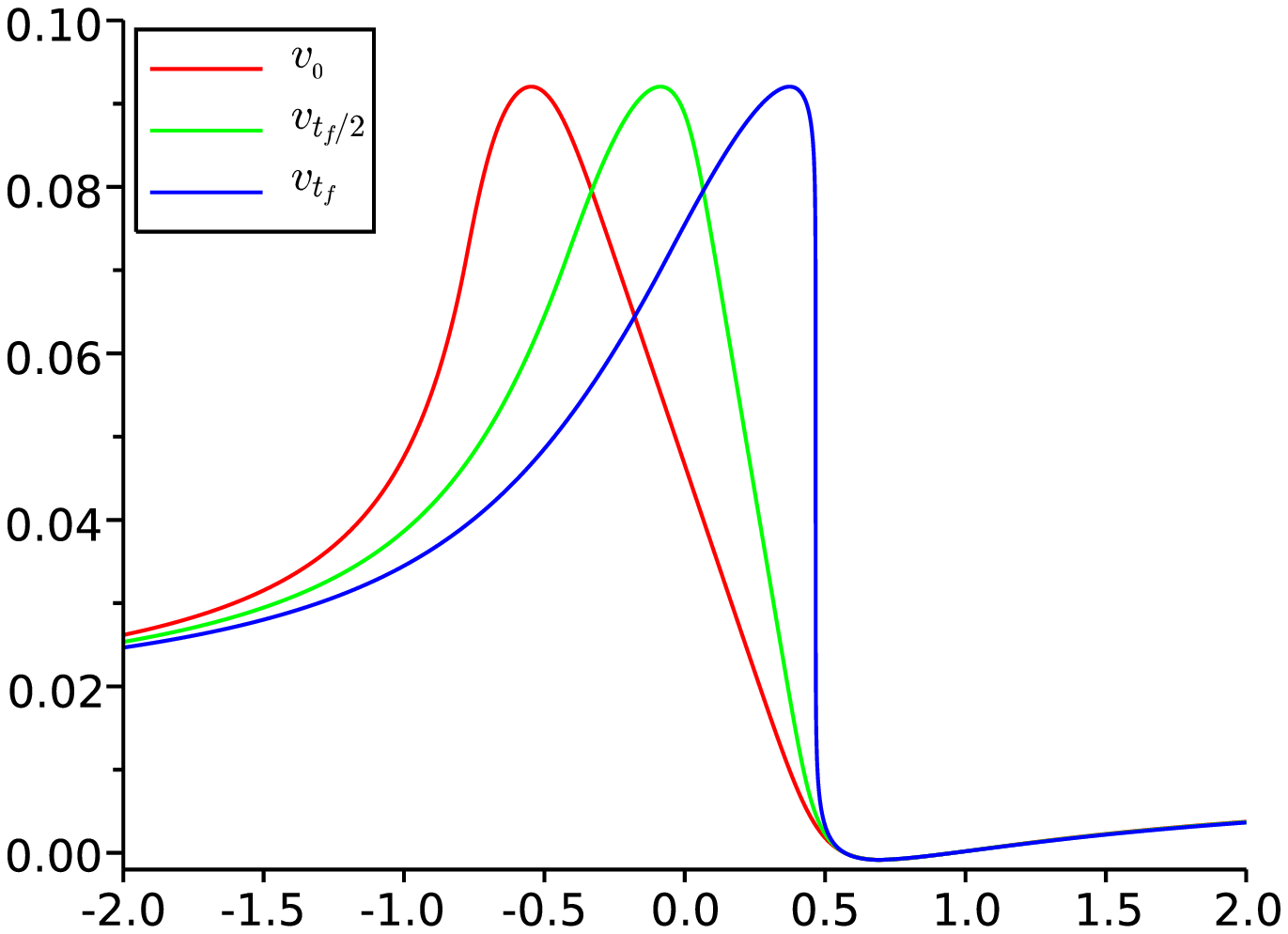}\\
        \vspace{-0.73cm} \strut
        \caption{Burgers velocity at initial,\\
                 \hspace*{1.2cm}half-time and final times,\\ 
                 \hspace*{1.2cm}time window $\,t_\ff=10s$}   
        \end{minipage}}
    \hspace*{0.8cm}
{%
      \begin{minipage}{0.4\textwidth}\vspace{0.17cm}\hspace*{-0.6cm}
        \includegraphics[width=7.8cm,height=5.7cm]{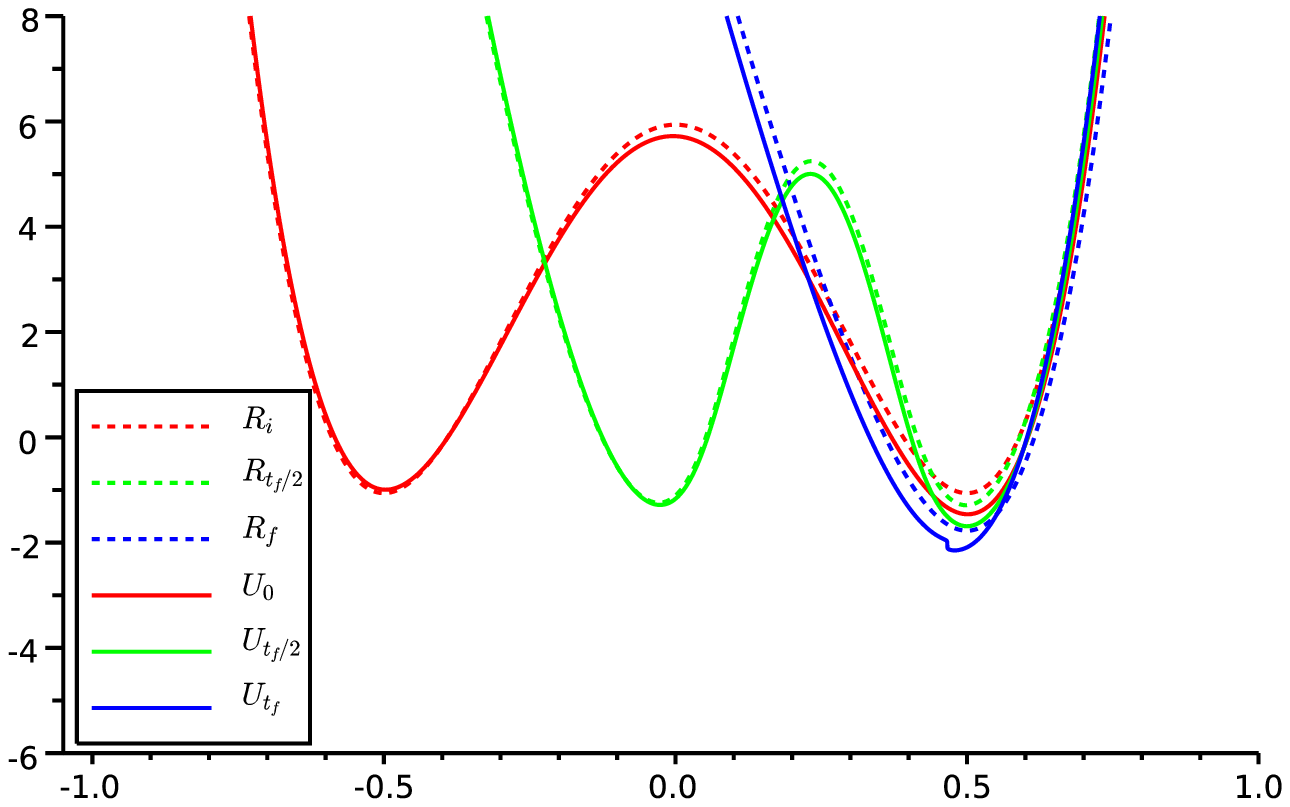}\\
        \vspace{-0.7cm} \strut
        \caption{\ initial, half-time and final potentials\\
                 \hspace*{1.2cm}initial, half-time and final controls,\\
                 \hspace*{1.2cm}time window $\,t_\ff=10s$}   
        \end{minipage}}
    \hspace*{7pt}
\end{center}
\end{figure}

\noindent (usually one takes $\,M\,$ equal or proportional to the unit matrix). 
\,The optimal assignment $\,\bm q_n\mapsto\bm 
x_{\pi(n)}\,$ gives a discrete approximation to the optimal Lagrangian map 
$\,\bm x_\ii\mapsto\bm x_\ff(\bm x_\ii)$. \,The optimal permutation
$\,\pi\,$ may be searched for employing a version of the Auction Algorithm
\citep{Berts}, see also Sec.\,4 of \citep{BFHLMMS}. In our one-dimensional
simulation, we took $\,N=10^5$. \,As an independent check of
the method, we also performed the two-dimensional simulations 
with the factor (\ref{ydensity}) included in the density, confirming 
the (approximately) factorized form of the resulting optimal assignment. 
Fig.\,3 compiles the results for the three 
methods of computation of the Lagrangian minimizer $\,x_\ii\to x_\ff(x_\ii)\,$ 
(the broken curve $\,x_\ii\mapsto x_\ii+\alpha/4\,$ is its exact asymptote
at $\,\pm\infty$) \,and of its derivative. The results agree well in the 
common domains. The green thick dots represent the assignment obtained with 
the Auction Algorithm. 
\vskip 0.2cm

Given the optimal Lagrangian map $\,x_\ff(x_\ii)$, \,we put the 
dynamics into it by interpolation defining
\qq
x(t;x_\ii)\,=\,\frac{{t_\ff-t}}{{t_\ff}}\,x_\ii\,+\,\frac{t}{t_\ff}
\,x_\ff(x_\ii)\,.
\qqq 
The corresponding Burgers velocities 
\qq
v(t,x)\ =\ \mu\,\nabla\Psi(t,x)\,=\,\frac{1}{{t_\ff}}\,
(x_\ff(x_\ii)-x_\ii)\Big|_{x(t;x_\ii)=x}
\qqq
are plotted in Fig.\,4 for $\,t_\ff=1\hspace{0.01cm}s\,$ as function of $\,x\,$ 
at times $\,t=0$, $\,t=t_\ff/2$, \,and $\,t=t_\ff$. \,As we see, the evolution 
of the Burgers velocities that displace the initial distribution to the 
final one over time $\,t_\ff\,$ describes a nascent shock. \,The corresponding 
dynamic potentials 
$\,R_t=-k_BT\,\ln\rho_t\,$ are drawn on Fig.\,5 which also plots the 
time-dependent optimal controls $U_t$. \,Note that the initial control 
$\,U_{0}\,$ is different from $\,R_\ii=R_{0}$, \,rearranging the symmetric 
wells by making the right one deeper. \,The half-time control 
$\,U_{t_\ff/2}\,$ moves the left metastable well further to the right. In 
the final control $\,U_{t_\ff}$, \,the left well disappears altogether. 
On the other hand, \,for the time interval 10 times longer, controls 
$\,U_\ii\,$ become close to dynamic potentials $\,R_t$, \,see Figs.\,6 
and 7 \,($U_t\,$ would coincide with $\,R_t\,$ for 
an infinitely slow process).     
\vskip 0.3cm

The quadratic cost function corresponding to the optimal Lagrangian map 
$\,x_\ii\mapsto x_\ff(x_\ii)\,$
\qq
\frac{_1}{^{k_BT}}\,\CK_{min}\,\approx\,1.996448\hspace{0.04cm}s 
\qqq
(the Auction Algorithm produced a value lower by $\,0.07\%$, \,giving
an idea about the accuracy of our calculations).
The minimal average heat release during the process
with duration $\,t_\ff=1\hspace{0.01cm}s\,$ is
\qq
\big\langle\,Q\,\big\rangle^{t_\ff=1s}_{min}\,\,\,\approx\,
\,(1.996448/1\,+\,0.7431204)\,k_BT\,\approx\,2.7395684\,k_BT\,\ 
\qqq
whereas for the 10 times longer process
\qq
\big\langle\,Q\,\big\rangle^{t_\ff=10\hspace{0.01cm}s}_{min}\,\,\,\approx\,\,
(1.996448/10\,+\,0.7431204)\,k_BT
\,\approx\,0.9427652\,k_BT\,.
\qqq
The average heat release exceeds the Landauer bound 
$\,(\ln{\hspace{-0.04cm}2})\hspace{0.04cm}k_BT\approx0.6931472
\,k_BT\,$ almost 4 times in the first case and by about 36\% in 
the second one.

\nsection{Extension to non-conservative forces}
\label{sec:NCF}

\noindent The overdamped Langevin dynamics (\ref{Lan}) has the drift
given by the gradient of a potential. In the presence of non-conservative 
forces, it should be modified to the stochastic equation
\qq
d\bm x\ =\ {M}\hspace{0.02cm}\big(-\bm\nabla U(t,\bm x)
+\bm f(t,\bm x)\big)\,dt\,+\,d\bm\zeta(t)\,,
\label{Lan1}
\qqq
where $\,\bm f\,$ represents such forces. We shall keep the noise as 
before assuming that the environment is still thermal and
the Einstein relation (\ref{ER}) holds. Eq.\,(\ref{Lan1}) 
defines again a Markov diffusion process $\,\bm x(t)\,$ with generator 
\qq
L_t\ =\ (-(\bm\nabla U_t)+\bm f_t)\cdot{M}\,\bm\nabla
+k_BT\hspace{0.07cm}\bm\nabla\cdot{M}\,\bm\nabla\,.
\label{Lt1}
\qqq
The Fokker-Planck equation (\ref{rhoev}) still takes the form
of the advection equation (\ref{rhoadv}) in the mean local
velocity field (\ref{limpr}) that becomes
\qq
\bm v(t,\bm x)\,&=&\,-\hspace{0.02cm}M\,\big({\bm\nabla}U+k_BT\rho^{-1}
{\bm\nabla}\rho\big)(t,\bm x)\ =\ -\hspace{0.02cm}M\,(\bm\nabla U
-\bm\nabla R-\bm f)(t,\bm x)\,.
\label{mlv1}
\qqq
The fluctuating heat release is now given by a generalization 
of formula (\ref{DQ}):
\qq
Q\ =\ \int_{0}^{t_\ff}\hspace{-0.05cm}\big(-{\bm\nabla}
U(t,\bm x(t))+\bm f(t,\bm x(t))\big)\cdot\circ\,d\bm x(t)\,.
\label{DQ1}
\qqq
with the expectation value
\qq
\big\langle\,Q\,\big\rangle\ =\ 
\int_{0}^{t_\ff}\hspace{-0.1cm}dt\int
(\bm\nabla U-\bm\nabla R-\bm f)(t,\bm x)\cdot{M}\,({\bm\nabla}U-\bm f)
(t,\bm x)\,\,\rho(t,\bm x)\,\,dx\,.
\label{avQ11}
\qqq
On the other hand, the change of the entropy of the system
takes in the presence of force $\,\bm f\,$ the form
\qq
\Delta S_{sys}\,&=&\,-\,k_B\int_{0}^{t_\ff}\hspace{-0.1cm}dt
\int\ln(\rho_t(\bm x))\,\,(L_t^\dagger\rho_t)(\bm x)\,\,dx\cr\cr
&=&\,-\,\frac{1}{T}\int_{0}^{t_\ff}\hspace{-0.1cm}dt
\int(\bm\nabla R)(t,\bm x)\cdot M\,
(\bm\nabla U-\bm\nabla R-\bm f)(t,\bm x)\,\,\rho(t,\bm x)\,\,dx\,.
\qqq 
Defining the entropy change in the environment $\,\Delta S_{env}\,$
by the thermodynamic relation (\ref{DSenv}), we infer that
the total entropy production in the time interval $\,[0,t_\ff]\,$ is again 
given by the right hand side of Eq.\,(\ref{DStot}):
\qq
\Delta S_{tot}\ =\ \Delta S_{sys}+\Delta S_{env}
\ =\ \frac{1}{T}\int_{0}^{t_\ff}\hspace{-0.1cm}dt\int
\big(\bm v\cdot M^{-1}\bm v\big)(t,\bm x)\,\,\rho(t,\bm x)\,\,dx\,,
\label{DStot1}
\qqq
see refs. \citep{CCJ} or \citep{FR} for the interpretation of 
$\,\Delta S_{tot}\,$ as a relative entropy of the processes with
direct and time-reversed protocols. Eq.\,(\ref{DStot1}) implies 
that the Second Law inequalities (\ref{2ndLaw}) and (\ref{bdforQ})
still hold in the presence of non-conservative forces. Recall, that 
in the previous discussion, we minimized the right hand side of the 
above expression for $\,\Delta S_{tot}\,$ over all velocity fields 
with densities $\,\rho\,$ evolving by the advection equation 
(\ref{rhoadv}) between the fixed initial and final ones. Hence the 
bounds (\ref{R2ndLaw}) and (\ref{RbdforQ}) providing a finite-time 
refinements of the Second Law still hold in the presence of 
non-conservative forces. They are saturated, nevertheless, by the 
dynamics with a conservative force that was constructed before.  
\vskip 0.2cm

The work performed on the system, admitting the possibility of potential 
jumps at the end-points of the time interval, \,is defined now by the 
expression
\qq
W\,&=&\,U(0,\bm x(0)-U_\ii(\bm x(0))\,
+\int_{0}^{t_\ff}\hspace{-0.1cm}
\partial_tU(t,\bm x(t))\hspace{0.08cm}dt\,+\,U_\ff(\bm x(t_\ff))
-U(t_\ff,\bm x(t_\ff))\cr
&&\,+\,\int_{0}^{t_\ff}\hspace{-0.1cm}
\bm f(t,\bm x(t))\cdot\circ\,d\bm x(t)\,\ =\ \Delta U\,+\,Q\,,
\qqq
where $\,\Delta U\,$ is given by Eq.\,(\ref{DU1}). \,Hence 
the minimization of the average work $\,\big\langle\,W\,\big\rangle\,$
for fixed $\,U_\ii$, $\,\rho_\ii$, \,and $\,U_\ff\,$ may be performed
as in Sec.\,\ref{sec:R2ndLaw}. Consequently, the inequality (\ref{minW})
still holds in the presence of non-conservative forces, but it is saturated 
by the same protocol as before, with a conservative force.

\nsection{Conclusions}
\label{sec:conc}

\noindent We have established an exact lower bound (\ref{R2ndLaw}) for 
the total entropy production in the overdamped Langevin dynamics with 
thermal noise interpolating in a fixed time window between given 
statistical states with smooth positive probability densities. The bound,
realizing a refinement of the Second Law (\ref{2ndLaw})
of Stochastic Thermodynamics, is valid in the presence of conservative 
or non-conservative driving forces. It is inversely proportional to the 
length of the time window and to the temperature. The proportionality 
constant is given by the minimum of the quadratic cost function 
(\ref{cost}) over all maps transporting the initial probability 
distribution to the final one. The minimal 
entropy production occurs for the process driven by a conservative 
force with a time-dependent potential expressed by solutions of 
the inviscid Burgers equation related to the optimal Monge-Kantorovich 
mass transport and of the accompanying advection equation for densities.
The refined Second Law (\ref{R2ndLaw}) induced the optimal lower bounds 
(\ref{RbdforQ}) and (\ref{minW}) for, respectively, the average heat
release and average performed work. The general theory was illustrated 
on the example of a Gaussian Langevin process and on a model describing 
a mesoscopic particle manipulated by optical tweezers with a memory erasure 
dynamics of the type discussed in \citep{DLutz}
as a toy model for Thermodynamics of Computation \citep{Bennett}. The system 
was recently studied experimentally \citep{Berut} and we plan to use the 
outcome of the numerical analysis of our model to suggest an improvement 
of the experimental protocol in order to lower the average heat release
in the process.  
\vskip 0.1cm

Our results should have a simple extension to the case with 
limiting states given by probability measures without smooth densities. 
Such an extension would involve viscosity solutions of the inviscid 
Burgers equation admitting shocks. Applications to cyclic processes 
(e.g. to models of molecular motors and to optimization of their 
efficiency is among natural directions of further research, see 
\citep{SS1} and the references therein. A more difficult problem requiring
limiting arguments is an extension of the above results to the case 
of underdamped Langevin dynamics. A related discussion of work minimization 
in the Gaussian case may be found in \citep{GMSS}. Another step in that 
direction was taken recently in \citep{AMMMG1}. In general, the question 
about the minimal entropy production in finite time processes between 
fixed states makes sense for more general modelizations of 
non-equilibrium dynamics, e.g. for the ones involving thermostats. 
It is certainly worth studying in such contexts. Other optimization 
problems of deterministic or stochastic nature related to fluctuation 
relations in non-equilibrium statistical mechanics may also be  interesting 
\citep{AMMMG}. The optimization techniques \citep{Bellman,FlemSon} developed 
largely with an eye on other cost functions, seem to find this way new 
important applications. 

\end{document}